\documentclass[12pt]{iopart}

\usepackage{subfig}

\usepackage[LGR,T1]{fontenc}
\usepackage[utf8]{inputenc}
\usepackage{color}
\usepackage{wrapfig}
\usepackage[separate-uncertainty = true,multi-part-units=single]{siunitx}
\usepackage{xfrac}
\usepackage{textgreek}
\usepackage{textcomp}
\usepackage{amstext}
\usepackage{graphicx}
\graphicspath{{./}{figs/}}
\usepackage{subscript}
\usepackage{todonotes}
\usepackage{booktabs}

\expandafter\let\csname equation*\endcsname\relax
\expandafter\let\csname endequation*\endcsname\relax
\usepackage{amsmath}
\usepackage{newunicodechar}
\newcommand{\etal}{\emph{et al.\ }}
\newcommand{\ExB}[0]{{ $ \vec{E} \times \vec{B}$ }}

\usepackage[backend=biber,style=numeric, sorting=none]{biblatex}
\AtEveryBibitem{
  \clearfield{note}
  \clearfield{issn}
  \clearfield{url}
  \clearfield{urlday}  
  \clearfield{urlmonth}  
  \clearfield{urlyear}  
  \clearfield{doi}  
}

\begin{document}

\title{Azimuthal ion movement in HiPIMS plasmas - Part I: velocity distribution function}

\author{S Thiemann-Monjé, J Held\footnote{current affiliation: University of Minnesota, Minneapolis, USA}, S Schüttler\footnote{current affiliation: Plasma Interface Physics, Ruhr University Bochum, Bochum, Germany}, A von Keudell, V Schulz-von der Gathen}

\address{Experimental Physics II, Ruhr University Bochum, Germany}
\ead{Achim.vonKeudell@ruhr-uni-bochum.de}
\begin{abstract}
Magnetron sputtering discharges feature complex magnetic field configurations to confine the electrons close to the cathode surface. This magnetic field configuration gives rise to a strong electron drift in azimuthal direction, with typical drift velocities on the order of \SI{100}{\kilo\meter\per\second}. In high power impulse magnetron sputtering (HiPIMS) plasmas, the ions have also been observed to follow the movement of electrons with velocities of a few \si{\kilo\meter\per\second}, despite being unmagnetized. In this work, we report on measurements of the azimuthal ion velocity using spatially resolved optical emission spectroscopy, allowing for a more direct measurement compared to experiments performed using mass spectrometry. The azimuthal ion velocities increase with target distance, peaking at about \SI{1.55}{\kilo\meter\per\second} for argon ions and \SI{1.25}{\kilo\meter\per\second} for titanium ions. Titanium neutrals are also found to follow the azimuthal ion movement which is explained with resonant charge exchange collisions. The experiments are then compared to a simple test-particle simulation of the titanium ion movement, yielding good agreement to the experiments when only considering the momentum transfer from electrons to ions via Coulomb collisions as the only source of acceleration in azimuthal direction. Based on these results, we propose this momentum transfer as the primary source for ion acceleration in azimuthal direction.

\end{abstract}


\maketitle

\section{Introduction}
\label{sec:intro}
Magnetron sputtering processes are widely used in industry for thin film deposition \cite{glocker_handbook_1995}. Traditionally, magnetron sputtering discharges are driven with continuous voltage (DCMS). However, in recent years, high power impulse magnetron sputtering (HiPIMS) has become more and more relevant. HiPIMS plasmas are excited with short high voltage pulses, leading to high current densities and peak pulse powers. At typical duty cycles of a few percent at most, the time-averaged power is kept low to prevent target melting. 

The high pulse power in HiPIMS discharges results in plasma densities ranging from $10^{19}$ m$^{-3}$ to $10^{20}$ m$^{-3}$ \cite{hecimovic_probing_2017, held_electron_2020, gudmundsson_spatial_2002} and ionization degrees of the sputtered particles of up to 90\% \cite{biskup_influence_2018,kouznetsov_novel_1999,poolcharuansin_ionized_2012,bohlmark_ionization_2005} leading to superior coating qualities \cite{alami_ion-assisted_2005,sarakinos_process_2007}. The main drawback of HiPIMS discharges are the often observed lower deposition rates compared to DCMS discharges operated at similar average powers \cite{horwat_spatial_2008,anders_deposition_2010}.

The geometry for magnetron sputtering discharges is often cylindrical symmetric with a circular cathode, the so-called target. Two concentric ring magnets placed behind this target are forming arch-shaped magnetic field lines in radial direction, trapping the electrons to the region close to the target. This magnetic trap configuration then leads to a torus-shaped plasma. Consequently, sputtering is mostly taking place in the ring-shaped area below the plasma torus forming an equally shaped erosion area, the so-called racetrack. 

Above the racetrack area, the magnetic field is parallel to the target surface while the electric field vector points towards the target \cite{rauch_plasma_2012}. On one hand, this electric field pulls ionized sputtered particles back towards the target, hindering them from reaching the substrate and lowering the deposition rate of HiPIMS discharges \cite{brenning_understanding_2012}. On the other hand, the crossed electric and magnetic field configuration induces a significant electron \ExB drift. Additionally, curvature and diamagnetic drifts are also present, adding up to azimuthal electron drift velocities in the order of \SI{100}{\kilo\meter\per\second} in the case of HiPIMS \cite{kruger_reconstruction_2018,rauch_estimating_2013}.


The ion movement in axial direction has been studied by several authors \cite{hnilica_revisiting_2020,hnilica_revisiting_2020-1} as being dictated by the electric field \cite{mishra_evolution_2010}, collisions \cite{held_velocity_2020} and the sputtering process. The ion movement in azimuthal direction has been studied by Lundin et al. \cite{lundin_cross-field_2008}, who placed a mass spectrometer at positions tangential to the racetrack of an HiPIMS-discharge with a titanium target to capture ions leaving the target region tangentially either in the direction of the \ExB movement or against it. They found the energy of fast titanium ions to be larger by about \SI{10}{\electronvolt} (or about \SI{2.5}{\kilo\meter\per\second}) in the direction of the \ExB movement. From these measurements, performed outside the magnetic trap, they concluded that the ions inside the magnetized region must be moving along the plasma torus, in the azimuthal direction of the discharge.

Since the ions in magnetron sputtering discharges are unmagnetized \cite{kruger_reconstruction_2018}, an $\vec{E} \times \vec{B}$ drift of ions can be excluded as the explanation of the observed movement. Lundin \etal  proposed momentum transfer from the drifting electrons onto the ions as the reason for the observed phenomenon \cite{lundin_cross-field_2008}. They speculated that a modified two-stream instability is excited by the difference in drift velocity between electrons and ions. The resulting azimuthal electric field can then accelerate the ions, slowly dragging them along with the electron drift. Simple estimations showed that such a force from electrons on ions mediated by an instability might indeed explain the observed behavior. 

Later, Poolcharuansin \etal  repeated the experiment, using a retarding field analyzer instead of a mass spectrometer \cite{poolcharuansin_more_2012}. They also found a difference of roughly \SI{10}{\electronvolt} or \SI{2.5}{\kilo\meter\per\second} for ions leaving the magnetic trap region tangentially in or against the \ExB direction. The authors combined their experiments with a fairly complex model, describing both the acceleration of ions in the azimuthal direction, as well as collisions and the conditions under which ions can even reach the detector, without being pulled back into the magnetic trap region by the electric field. From their model, the authors found support for the modified two-stream instability hypothesis proposed by Lundin \etal, explaining that ion-electron collisions alone would be insufficient to provide enough acceleration for the ions.

A different explanation for the same phenomenon was proposed by \textcite{panjan_asymmetric_2014} after performing similar experiments with an ion or electron collecting flat probe and a mass spectrometer, both again positioned tangentially to the target and outside the magnetic trap region. The authors found a correlation between the azimuthal ion movement and the appearance of spokes, another wave phenomenon present in magnetron sputtering discharges \cite{held_pattern_2020,hnilica_effect_2018, hecimovic_spokes_2018,kozyrev_optical_2011,ehiasarian_high_2012, anders_drifting_2012}. Spokes are known to cause plasma potential fluctuations and, thus, induce an asymmetric electric field \cite{panjan_plasma_2017,lockwood_estrin_triple_2017,held_electron_2020,held_spoke-resolved_2022}, which is expected to influence the ion movement, both in axial as well as in azimuthal direction \cite{held_electron_2020, anders_drifting_2013, panjan_plasma_2017, breilmann_fast_2017}.

All these prior measurements have in common that they observed only those ions that have left the magnetic trap region. Since most ions are expected to eventually return to the target surface, this group of ions leaving the magnetic trap region is not representative for the overall ion population inside the magnetic trap. Thus, gaining information about physical processes \emph{inside} the magnetic trap from such measurements is very challenging and prone to error.

The azimuthal movement of ions in magnetron plasmas is addressed in a two part series with part I addressing the velocity distribution functions of the ions inside the plasma and part II addressing the lateral deposition of species leaving the magnetic trap region\cite{schuttler_azimuthal_2023}. This paper constitutes part I, where we investigate the azimuthal ion movement using high-resolution optical emission spectroscopy. From the broadening and shifting of optical emission lines, we can directly determine the velocity distribution function of ions \emph{inside} the magnetic trap region, temporally and spatially resolved. The measurements are compared to a simple model, only considering the momentum transfer from electrons to ions via Coulomb collisions. We show that already such a momentum transfer via collisions alone can explain the observed ion velocities, without the need to consider wave phenomena, which do not seem to play a dominant role.

\section{Experimental setup}
\label{sec:setup}

\subsection{Chamber and discharge}
A cylindrical vacuum chamber with a diameter of 25 cm and a height of 40 cm was used for the experiment. It was pumped to a base pressure of $4 \times 10^{-6}$ Pa. Argon was used as working gas at a pressure of 0.5 Pa. A planar 2" magnetron (Thin Film Consulting IX2U) in combination with a TRUMPF Hüttinger power supply (TruPlasma Highpulse 4002) was used to drive the plasma discharges. 

The discharge was monitored by current and voltage measurements with commercial probes (Tektronix TCP A400, Tektronix P6015A) attached to the connection cable between the power supply and the magnetron assembly.  Discharge conditions were selected to be the same as in earlier publications \cite{held_velocity_2018,held_velocity_2020}. The applied voltage was -590 V with a repetition frequency of 40 Hz and a pulse length of 100 $\mu$s. Using titanium targets, these values result in peak currents of 50 A, peak target-area-normalized current densities of 2.5 Acm$^{-2}$ and peak power densities of 1.1 kWcm$^{-2}$.  The corresponding voltage and current waveforms can be found in a previous publication \cite{held_velocity_2018}. 

\begin{figure}
\centering
\includegraphics[width=75mm]{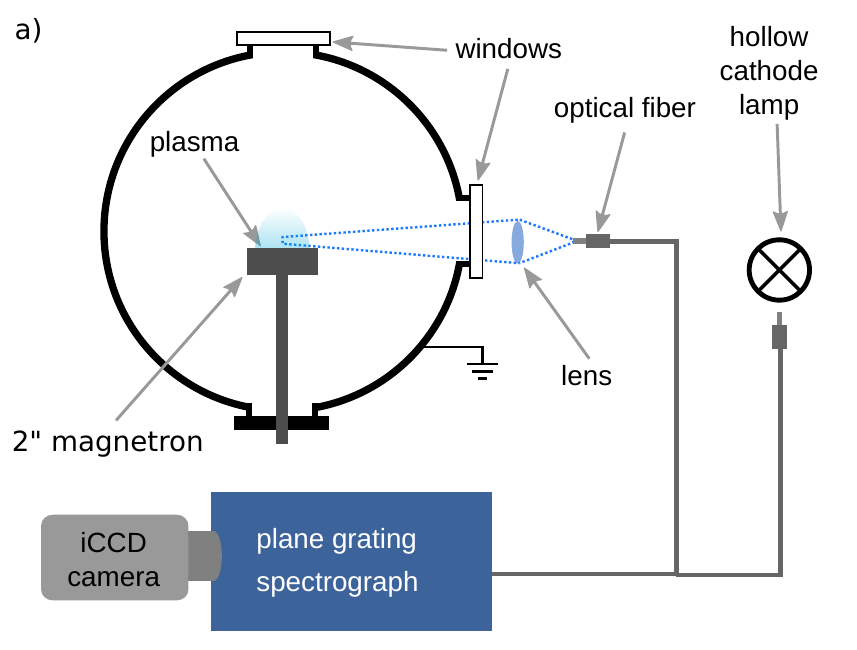} 
\includegraphics[width=75mm]{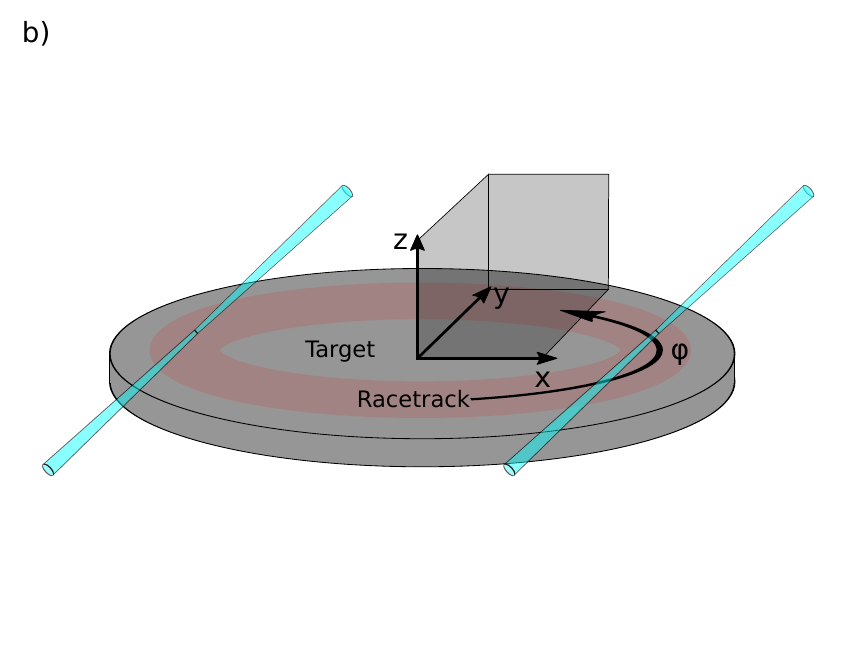}
\caption{a) Schematic of the high-resolution optical emission spectroscopy setup. Reproduced from \cite{held_velocity_2020}. b) A 3-dimensional depiction of the selected coordinate system and configuration of observation paths at x= $\pm$13.5 mm. }
\label{fig:setup}
\end{figure}


\subsection{High-resolution optical emission spectroscopy}

\begin{table}
    \centering
      \captionsetup{width=120mm}
    \caption{Energy levels and electron configurations of the selected optical emission lines \cite{saloman_energy_2012,whaling_argon_1995}  collected from the NIST atomic spectra database \cite{kramida_nist_2022}. The electron configuration is given in the notation first proposed by Russel \etal \cite{russell_report_1929}.}
    \label{tab:emission_lines}
    \begin{tabular}{c c c c}
         \hline \\
         Species & Ti\,II & Ar\,II &  Ti\,I  \\
         \hline
         Wavelength [nm] & 453.396   & 472.687 & 453.324 \\
         \\
         \textit{Upper level} &    &    & \\
         Energy [eV] &  3.97  &  19.76  & 3.58\\
         e$^-$-config. &  $3d^2(^3F)4p$  &  $3s^23p^4(^3P)4p$  & $3d^3(^4F)4p$\\
         \\
         \textit{Lower level} &    &    & \\
         Energy [eV] &  1.24 &  17.14  & 0.85\\
         e$^-$-config. & $3d^3$ &  $3s^23p^4(^3P)4s$  &$3d^3(^4F)4s$ \\
         \hline
    \end{tabular}
\end{table}

The setup for the high-resolution optical emission spectroscopy was adapted from \cite{held_velocity_2020} and is shown in figure \ref{fig:setup}a. The plasma is observed parallel to the target surface. A convex lens ($f$ = 150 mm) is used to collect the emitted light and couple it into an optical fibre (\O  = 800 $\mu$m). The distance between lens and fiber is adjusted to limit the field of view of the system to a narrow cone (figure \ref{fig:setup}b). The focal spot has a diameter of approximately 2 mm with the focal plane adjusted to the center of the target. The whole lens system is mounted on a movable stage and can be moved along the magnetron axis or parallel to the target surface ($z$ and $x$ direction). As illustrated in figure \ref{fig:setup}b, we define the $z$-axis of our coordinate system in target normal or axial direction and the $x$ and $y$-axis parallel to the target surface with the origin in the center of the target. Additionally, the coordinate $\varphi$ is used to describe the azimuthal ion movement. It points in the same direction as the \ExB drift.  

Measurements of the emission lines were performed with an intensified CCD-Camera (Andor iStar DH320T-25U-A3) attached to a 2 m plane grating spectrograph (Zeiss Jena PGS 2, 1300 lines/mm grating). All measurements were performed at the end of the discharge pulse by triggering the camera with a delay of $\Delta t = 90 \mu \text{s}$ to the plasma ignition. The gate width was set to \SI{10}{\micro\second} and the data were accumulated over 2000 plasma pulses, with the exception of the $x$-scan presented in figure \ref{fig:setup:x-var}b, where the last \SI{15}{\micro\second} of the pulse were measured and 500 accumulations were used, instead. By operating the spectrograph in the third diffraction order, an spectral resolution of \SI{1.5}{\pico\meter} pixel-to-pixel at the camera chip was achieved. To enable calibration of the wavelength axis for the measured spectra the emission from a hollow cathode lamp (HCL, Cathodeon 3UNX Ti) was measured simultaneously with the plasma emission, as indicated in figure \ref{fig:setup}a. Details about the used emission lines including the involved energy levels are displayed in table \ref{tab:emission_lines}.

The determination of the velocity distribution function (VDF) from the emission lines was performed as described in \cite{held_velocity_2018}. The method is based on the analysis of the two dominant line broadening mechanisms, Doppler broadening and instrumental broadening. As a first step, a Wiener deconvolution is used to remove the contribution of instrumental broadening to obtain an emission line profile only affected by Doppler broadening. Afterwards, the wavelength axis of the spectra is transformed into a velocity axis using the relation $v = c(\frac{\lambda}{\lambda_0} - 1)$, where $\lambda_0$ is the wavelength of the emission line in an unshifted state measured from the emission by the hollow cathode lamp.

\subsection{Probe measurements}

Probe measurements were performed above the racetrack position, in target distances of 6.3, 8.0 and 9.7 mm. The probe setup \cite{held_electron_2020} and results \cite{held_ionization_2023} are discussed in great detail in recent publications. Here, we only use the electron density and plasma potential obtained from those measurements to estimate the physical background and the corresponding forces acting on the ions.

\section{Results and Discussion}

\begin{figure}
\centering
\includegraphics[width=75mm]{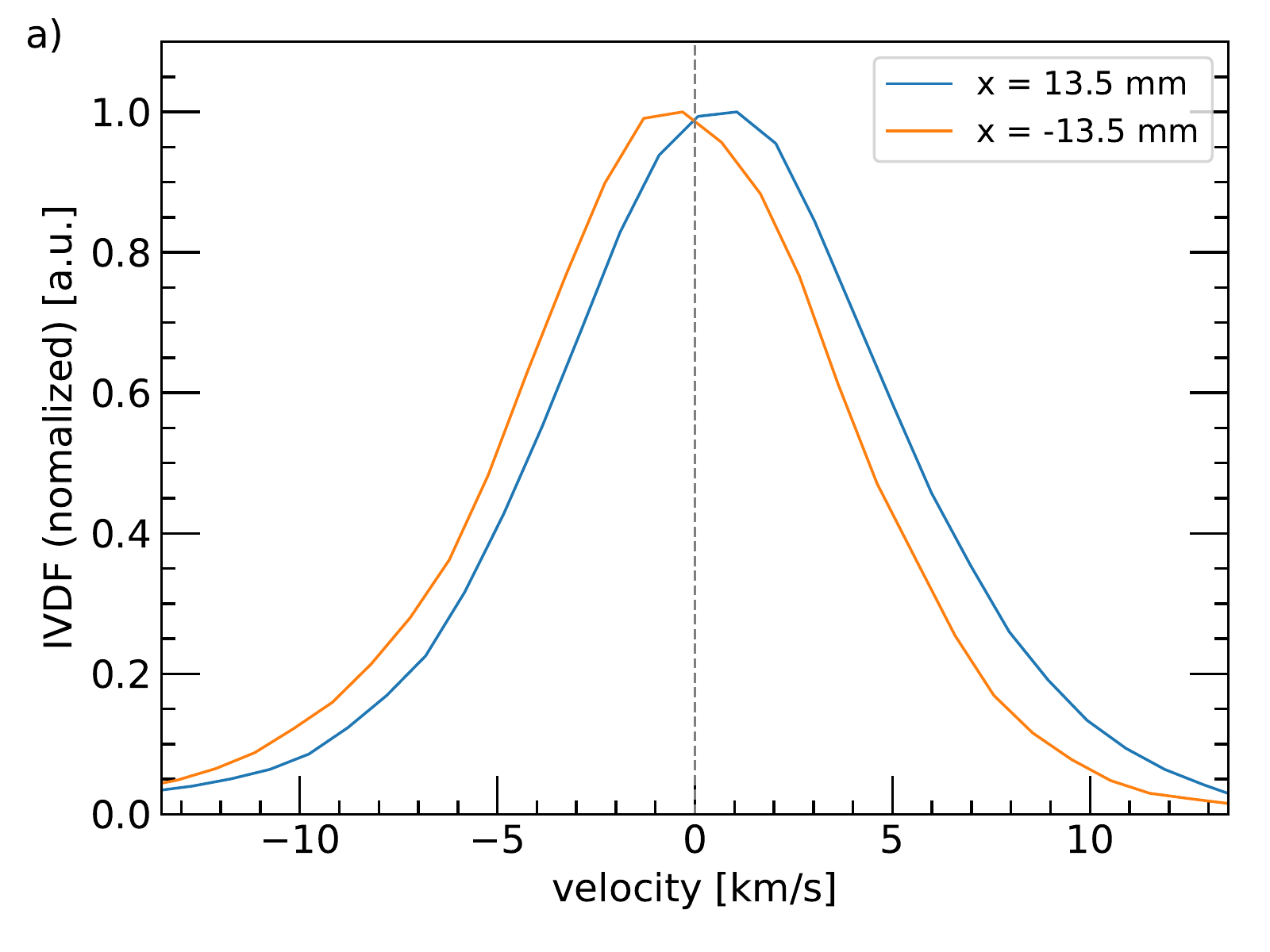} 
\includegraphics[width=75mm]{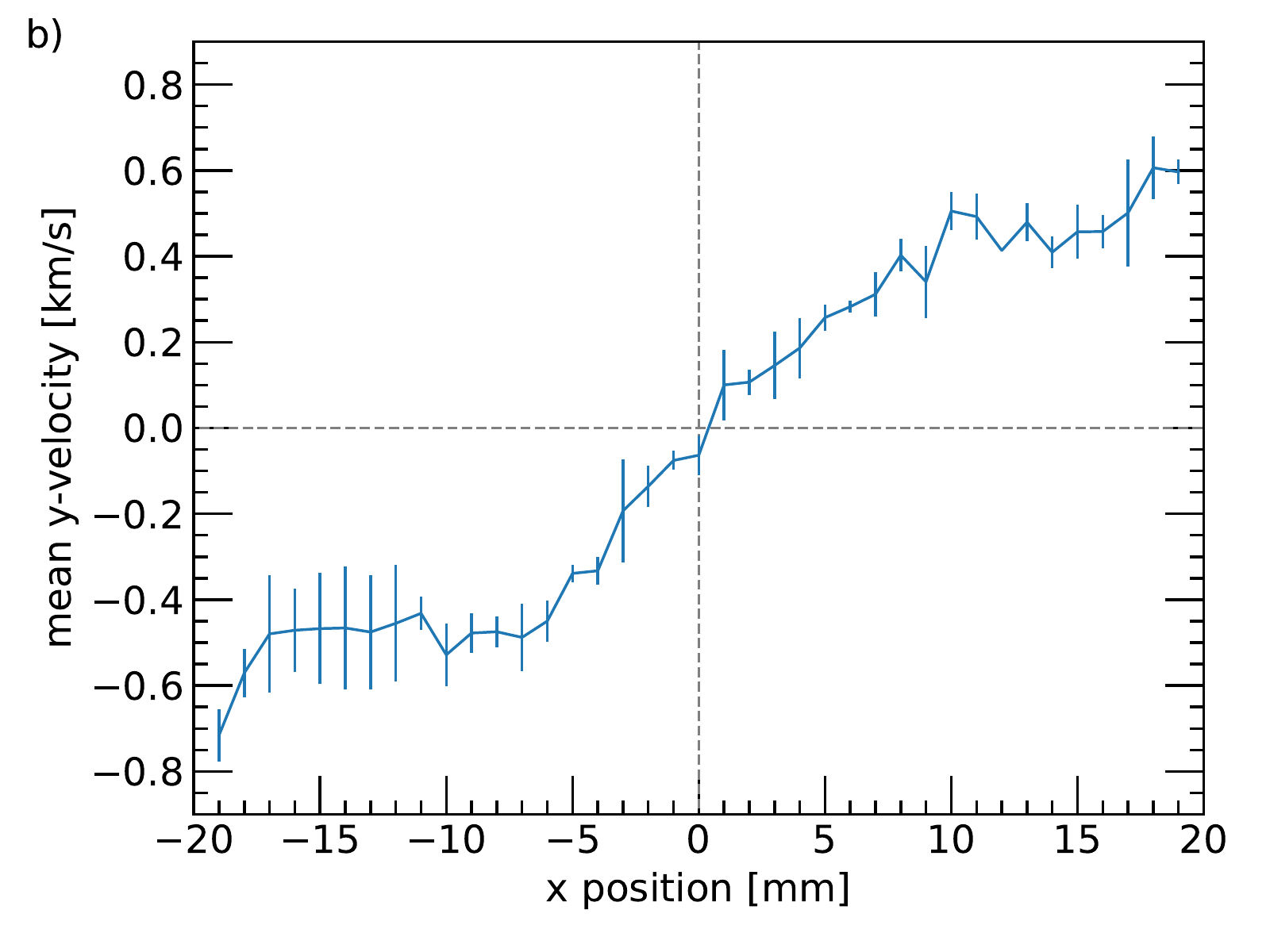}
\caption{a) Measured IVDF for the observation paths $x = -13.5$ mm and $x= 13.5$ mm (above the racetrack) at a distance of $z= 3$ mm.  \\
b) Measurements of the ion velocity $v_y$ for positions from $x = -20$ mm to $x = 20$ mm at a fixed target distance of $z= 3$ mm. }
\label{fig:setup:x-var}
\label{fig:setup:TiIIshift}
\end{figure}

Figure \ref{fig:setup:TiIIshift} a) shows an example of two obtained VDFs for titanium ions (Ti\,II). The optical system for both measurements was aligned to point in $y$-direction (compare figure \ref{fig:setup}b) at a fixed distance of $z = \SI{3}{\milli\meter}$. Intending a measurement above the racetrack on each side of the target, the measurement position in $x$-direction was selected to be $x = \pm\SI{13.5}{\milli\meter}$, as shown in figure \ref{fig:setup}b. As figure \ref{fig:setup:TiIIshift} demonstrates, a clear shift between the VDFs is observed, while their shapes remain the same. The VDF recorded at $x = \SI{13.5}{\milli\meter}$ is shifted to positive values by about \SI{0.5}{\kilo\meter\per\second}, indicating a mean particle movement away from the optical system (in positive $y$-direction). Since the VDF is symmetrical, except for its shift, the mean velocity is also calculated to be \SI{0.5}{\kilo\meter\per\second}. On the opposite side, at $x = -\SI{13.5}{\milli\meter}$, the VDF is shifted to negative values by the same amount, hence indicating a mean particle movement towards the optical system (in negative $y$-direction). In both cases, the movement follows the direction of the \ExB - drift, demonstrating that ions move along the racetrack in azimuthal direction together with the electrons - but much slower. 

Figure \ref{fig:setup:x-var} b) shows the mean titanium ion velocity in $y$-direction, calculated from the VDFs, for different $x$ positions, again at a fixed target distance of $z = 3 \text{ mm}$. Error bars indicate the standard deviation of three consecutive measurements. The displayed data show the expected change of $v_y$ due to the changing angle between the optical axis and the azimuthal particle movement. At $x = 0$, the azimuthal direction $\varphi$ is entirely perpendicular to the measurement direction $y$, so that the mean velocity in $y$ direction is $v_y = 0$. As such, correctly deducing the azimuthal velocity at each \emph{radial} position would require Abel-inversion of the line-of-sight integrated measurement data. Unfortunately, this would increase the noise of the measured data, leading to problems with the deconvolution used to obtain the VDF from the measured emission line profiles. However, figure \ref{fig:setup:x-var} shows constant values of $v_y = -0.5 \text{ km/s}$ and $v_y = 0.5 \text{ km/s} $ across the whole width of the racetrack region $ -16 \text{ mm} \leq x \leq -11 \text{ mm}$ and $ 11 \text{ mm} \leq x \leq 16 \text{ mm}$. This indicates that the particularly bright emission above the racetrack dominates over the contributions from all other radial positions, rendering $v_y \approx v_\varphi$ and allows us to use values measured at these positions as the azimuthal ion velocity. Consequently, all further measurements reported here were performed in the middle of the racetrack at  $x = \pm 13.5 \text{ mm}$ where the absolute value of the measured velocity represents the mean azimuthal velocity. All measurements were performed on both sides of the racetrack ( $x = 13.5 \text{ mm}$ and $x = -13.5 \text{ mm}$), and examined to ensure that the results are perfectly mirrored, i.e.\ positive velocities on one side exhibit the same magnitude of negative velocity on the other side. In this way, it was ensured that the measured velocities really represent the azimuthal movement of particles and are not distorted by any influence of other emission lines or a possible misalignment of the optical system. From here on, positive values for this azimuthal velocity indicate movement in \ExB direction.

\subsection{Azimuthal particle velocities}

\begin{figure}[h]
\centering
\includegraphics[width=75mm]{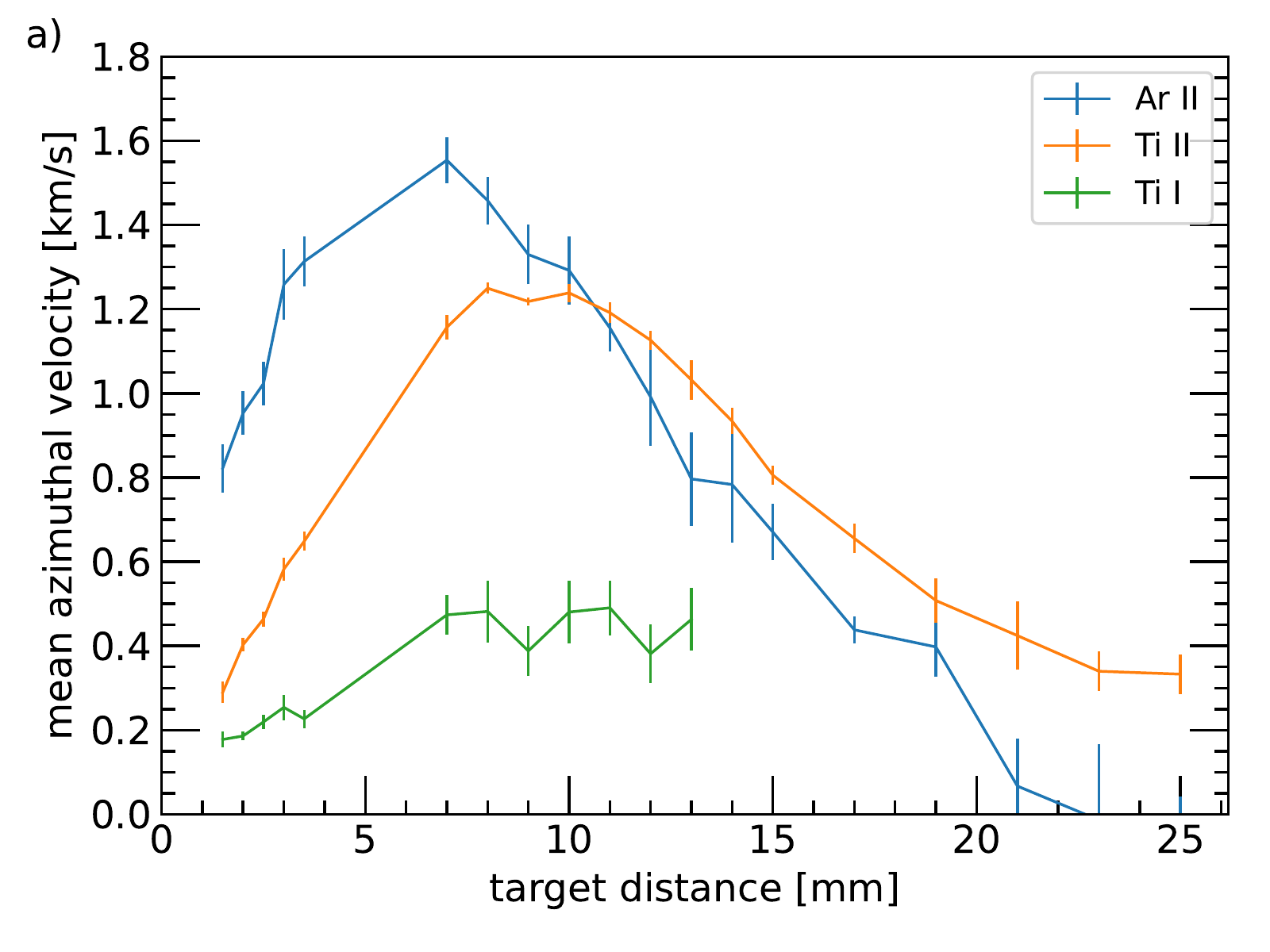}
\includegraphics[width=75mm]{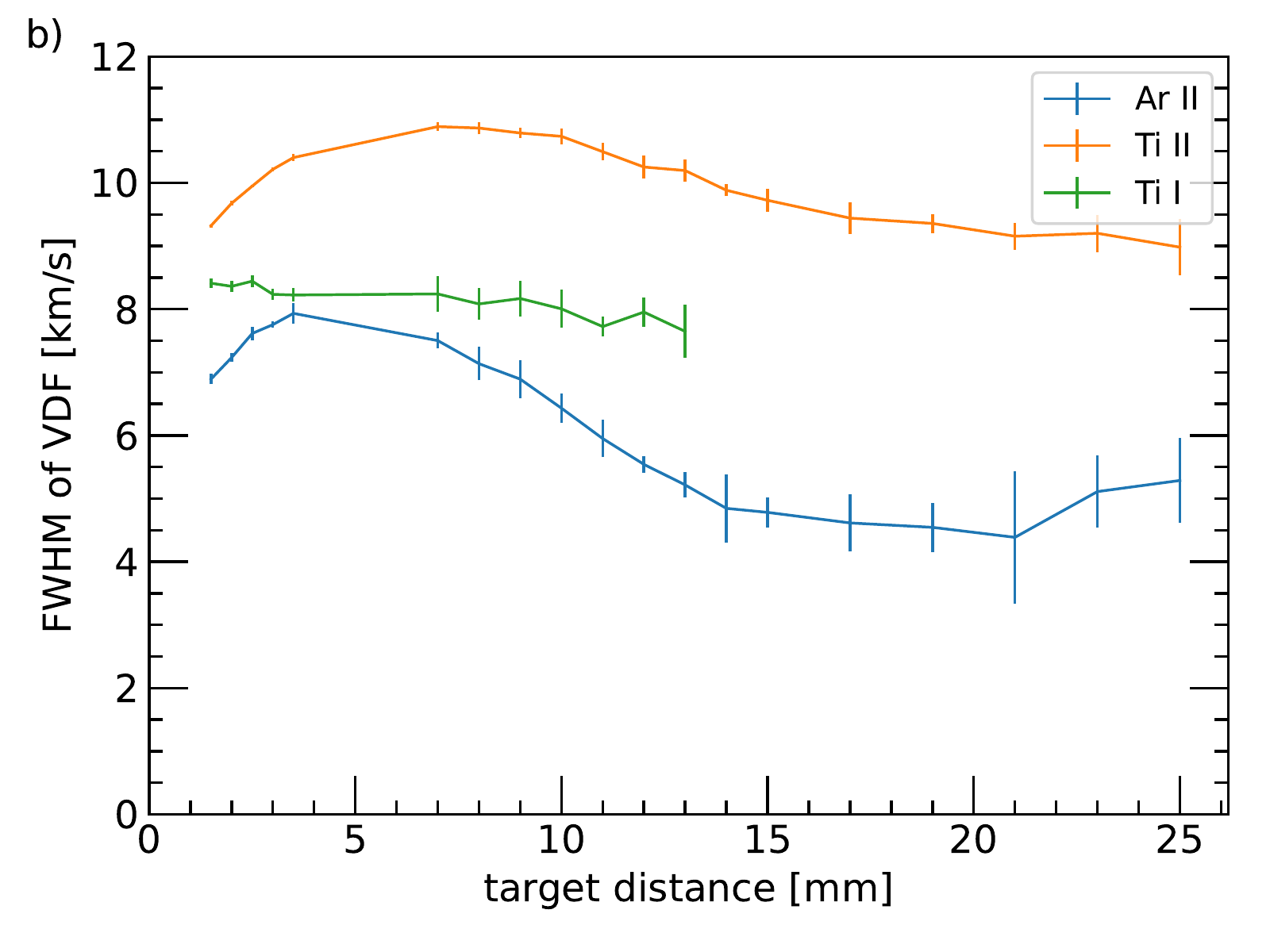}
\caption{a) Variation of the average azimuthal velocities for argon ions (Ar\,II), titanium ions (Ti\,II) and titanium neutrals (Ti\,I) with distance $z$ to the target surface. \\
b) Full width at half maximum (FWHM) of the measured velocity distribution functions. }
\label{fig:results:vrot-zvar}
\label{fig:results:fwhm-zvar}
\end{figure}

The dependence of the average velocity on the distance to the target surface is shown in figure \ref{fig:results:vrot-zvar}a. Considering argon ions (Ar\,II) first, we initially observe a steep increase with increasing target distance from about $v_{\varphi \text{ArII}} = \SI{0.8}{\kilo\meter\per\second}$ at $z = \SI{1.5}{\milli\meter}$ to a maximum of $v_{\varphi \text{ArII}} = \SI{1.55}{\kilo\meter\per\second}$ around $ z = \SI{7}{\milli\meter}$. The gap in the measurement data around  $z=\SI{5}{\milli\meter}$ is caused by the anode cover, blocking the field of view of the optical measurement system, as explained in a recent publication \cite{maas_synchronising_2021}. For higher values of $z$, $v_{\varphi \text{ArII}}$ begins to decrease with target distance and vanish at $z = \SI{23}{\milli\meter}$. A similar trend is observed for Ti\,II, with a smaller maximum velocity of $v_{\varphi \text{TiII}} =\SI{1.25}{\kilo\meter\per\second} $ peaking at a slightly larger target distance of $ z = \SI{10}{\milli\meter}$. 


Qualitatively, the observed trends in azimuthal ion velocity with varying target distance can be understood by considering the trajectories of the ions. For titanium, particles are created by sputtering at the target surface and are then ionized very close to the target, at $z < \SI{1}{\milli\meter}$, as we recently reported \cite{held_ionization_2023, held_transient_2021}. Particles entering the plasma from the target exhibit an initial azimuthal velocity of  $v_{\varphi} = 0$ and are subsequently accelerated inside the plasma. Thus, $v_{\varphi}$ increases with target distance for Ti\,II as particles travel through the plasma and are continuously accelerated in the azimuthal direction. The $v_{\varphi}$ decrease at larger target distances $z > \SI{10}{\milli\meter}$ can be explained by the lack of any additional acceleration force at these positions, independent of which physical process is actually causing the acceleration: waves are expected to be much weaker at this position \cite{held_spoke-resolved_2022, panjan_plasma_2017, held_electron_2020} and the electron density much lower, leading to less momentum transfer to the ions. Instead, the ions are only slowed down by collisions with the background gas, leading to smaller azimuthal velocities. On top of that, only the fastest ions can overcome the electric field and reach positions with $z > \SI{10}{\milli\meter}$. Because of their large velocity, such ions have crossed the dense plasma region close to the target very quickly, leaving them not much time to be accelerated in azimuthal direction. This effect will be explored in more detail in section \ref{sec:simu}.

The larger maximum azimuthal velocity for argon ions compared to titanium ions can simply be explained by their smaller mass: the effective force acting on the ions is  independent of the ion mass according to the reported explanations for the azimuthal ion movement found in the literature. As such, the ion acceleration is expected to scale with the ion mass $m$ as $m^{-1}$. This would lead to a difference in the maximum velocity of $m_{Ti}/m_{Ar} = 1.2$, which accounts for almost all of the observed differences in azimuthal velocity.

On top of the difference in maximum azimuthal velocity, figure \ref{fig:results:vrot-zvar}a also shows that the position of peak velocity is different for the two ion species: for titanium ions, the velocity peaks at $z = \SI{10}{\milli\meter}$, whereas the peak position is at about $z = \SI{7}{\milli\meter}$ for argon ions. This shift can likely be explained by the difference in location \emph{where} ionization occurs for the different ion species. Assuming strong working gas rarefaction, the direct vicinity of the target will be void of argon neutrals, and ions are rather created at some distance to the target surface. They are then accelerated towards the target by the electric field. Their average flow velocity towards the target leads to the observed shift in the position of maximum azimuthal velocity towards smaller $z$ values compared to titanium ions, which instead maintain a positive average flow velocity (away from the target) since they are created close to the target surface. The smaller argon ion velocity closer to the target surface, $z < \SI{7}{\milli\meter}$ might partly be caused by ion-ion collisions with the slower titanium ions and partly by mixing with argon ions created from neutrals that have outgassed from the target surface due to the working gas recycling \cite{anders_recycling_2012}.

For titanium neutrals, figure \ref{fig:results:vrot-zvar}a reveals a maximum azimuthal velocity of \SI{0.45}{\kilo\meter\per\second}, located somewhere around $z = \SI{10}{\milli\meter}$. Data for $z > \SI{13}{\milli\meter}$ could not be obtained, since the emission line used for the measurement was disturbed by titanium ion emission, which increases in relative intensity with the target distance. The observation that even neutral species have a considerable velocity in azimuthal direction is at first surprising, since neither of the two explanations proposed in the literature for the azimuthal acceleration - momentum transfer from the electrons or the electric field fluctuations caused by the spokes - applies to neutrals. We propose that the movement of neutrals is due to resonant charge exchange collisions with the titanium ions. Since titanium ions are expected to have a large density and the cross section for resonant charge exchange is very large for titanium ($\sigma_{cx} \approx \SI{2e-18}{\square\meter}$ \cite{smirnov_tables_2000}), the titanium neutrals are being dragged along with the azimuthal movement of ions. Assuming a titanium ion density of $n_{\mathrm{Ti}^+} = \SI{5e19}{\per\cubic\meter}$, the mean free path for resonant charge exchange is only $\lambda=(n_{\mathrm{Ti}^+} \sigma_{cx})^{-1} = \SI{10}{\milli\meter}$, demonstrating that our hypothesis is reasonable. This explanation is also in good agreement with the previously observed close coupling of titanium neutral and ion VDF in target normal direction \cite{held_velocity_2018}.

For argon neutrals, no azimuthal drift could be observed with our setup. However, previous work by Kanitz \etal did reveal a mean azimuthal velocity of about \SI{30}{\meter\per\second} for argon metastable atoms \cite{kanitz_two_2016}. This much lower azimuthal velocity can be explained by the smaller cross-section for resonant charge exchange for argon and the lower argon ion density in the target vicinity, leading to much less efficient momentum transfer from ions to neutrals.

Figure \ref{fig:results:vrot-zvar}b, shows the width (full width at half maximum - FWHM) of the VDF, as a measure of the average energy or effective temperature of the species. For titanium ions, the width of the VDF was already discussed in a recent publication \cite{held_velocity_2020}. There, we explained that titanium ions start their life as highly energetic sputtered particles following a Thompson energy distribution. The Thompson distribution is rather narrow (in terms of FWHM), but features a strongly populated high-energy tail. As particles move through the plasma they undergo Coulomb collisions with each other, leading to the relaxation of the VDF towards a Maxwell distribution. At the same average energy, the Maxwell distribution has a much larger FWHM, which is why the FWHM in figure \ref{fig:results:vrot-zvar}b increases with target distance until about $z = \SI{8}{\milli\meter}$ after which cooling by collisions with the background gas causes the VDF to become more narrow again. The maximum width observed here for Ti\,II is around \SI{11}{\kilo\meter\per\second}, which would correspond to a temperature of about \SI{10}{\eV} in case of a fully relaxed distribution. 

For argon ions, we generally find a much narrower VDF, with FWHMs between \SI{5}{\kilo\meter\per\second} and \SI{8}{\kilo\meter\per\second}. The reason for this smaller VDF width, which corresponds to a lower average energy, is that argon ions are created from argon neutrals, which are known to remain comparatively cold during the discharge pulse \cite{kanitz_two_2016, vitelaru_argon_2012}. As such, the newly ionized argon particles start out cold and are then heated up by Ohmic heating and by collisions with the energetic titanium ions, which leads them to acquire an effective temperature somewhere between the temperature of the cold argon neutrals and of the highly energetic titanium ions. The maximum VDF width observed for the argon ions (\SI{8}{\kilo\meter\per\second}) corresponds to a temperature of \SI{4.8}{\eV}.

For titanium neutrals, the VDF has an initial width of about \SI{8.5}{\kilo\meter\per\second}, corresponding to an unaltered Thompson distribution. For larger target distances, the VDF becomes slightly more narrow, presumably due to collisions with the working gas.


\subsection{Model for the ion movement}
\label{sec:results:model}

Following the qualitative description of the observed azimuthal velocities, we will now attempt to find a quantitative description. To this end, forces in both axial ($z$) as well as azimuthal ($\varphi$) direction need to be considered. The forces in $z$ direction determine the residence time of the particles within each volume element, which determines how much acceleration in $\varphi$ direction the passing species can accumulate. 

\subsubsection{Force in $z$ direction and electric field} 

The movement of ions in $z$-direction is mainly determined by the electric field in the magnetic trap region of the plasma, caused by the limited mobility of electrons across the magnetic field lines \cite{hagelaar_two-dimensional_2002}. As such, we need to find an estimation of the electric field configuration to describe the ion movement in this direction. 

\begin{figure}[h]
\centering
\includegraphics[width=75mm]{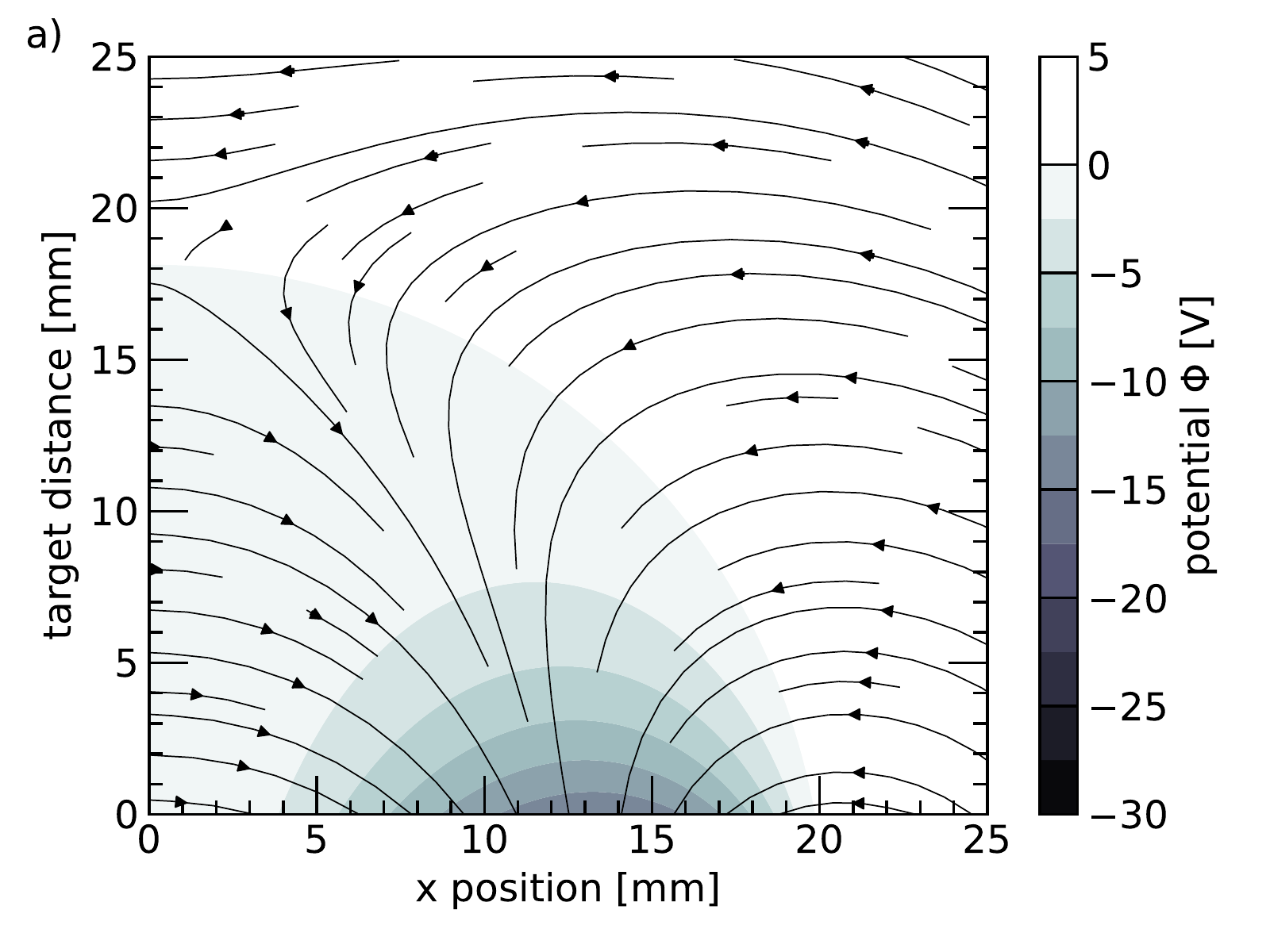}
\includegraphics[width=75mm]{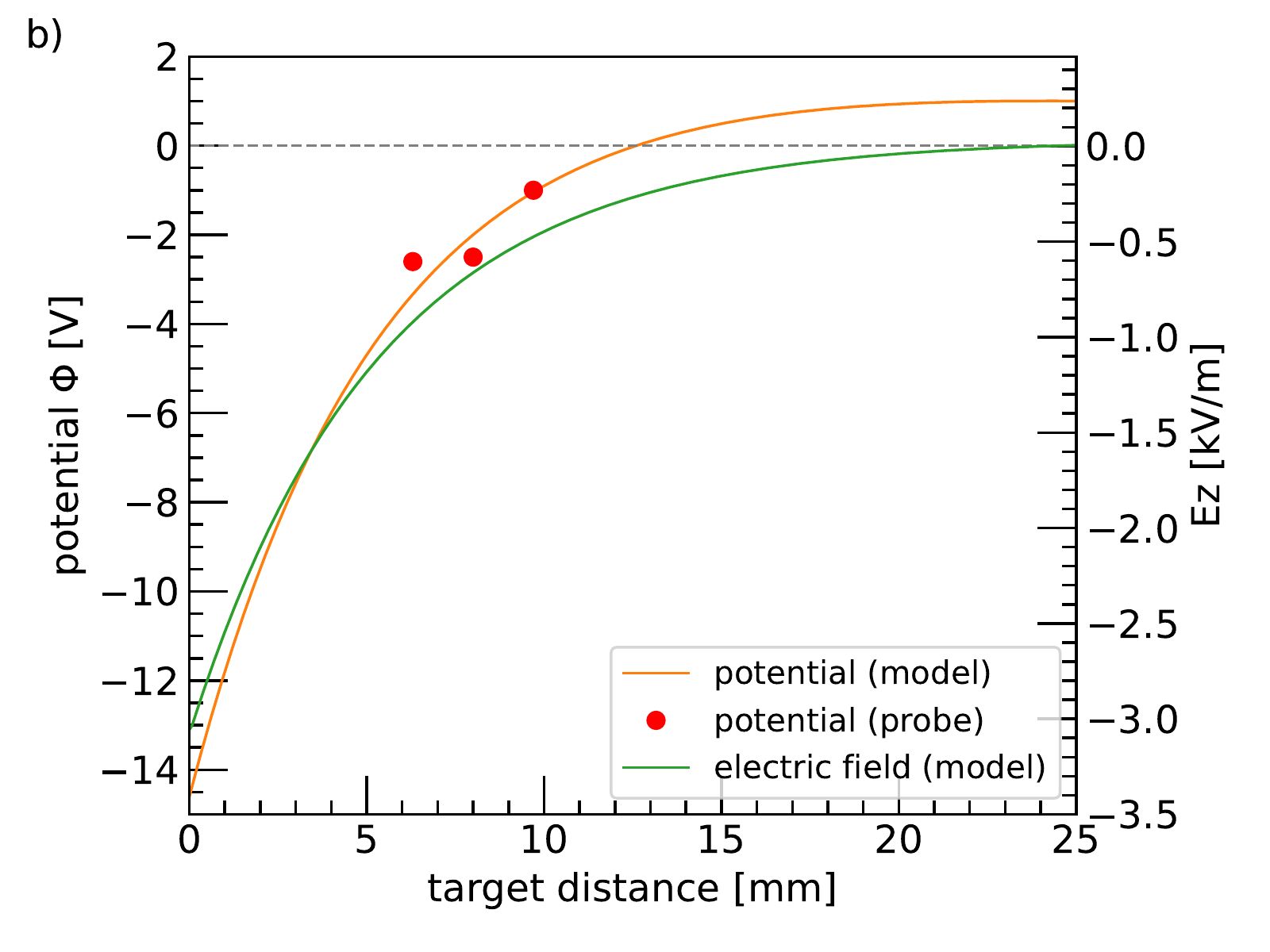}
\caption{a) Map of the plasma potential $\Phi$ deduced from the magnetic field topology assuming $\Phi \propto \Psi$. More intense colors indicate a more negative plasma potential. Arrows indicate the direction of the electric field. b) Plasma potential and electric field above the racetrack position, $r = \SI{13.5}{\milli\meter}$. }
\label{fig:plasma_pot}
\end{figure}

The electric field $\vec{E}$ is derived from the topology of the magnetic field $\vec{B}$ following a simple physical argument: due to the high mobility of electrons parallel to the magnetic field lines, any potential differences inside the magnetic trap region can only occur perpendicular to those. Therefore, the magnetic flux coordinates $\Psi$ introduced by \textcite{brinkmann_axisymmetric_2020} can be used to construct the topology of the electric potential $\Phi$ inside the magnetic trap region. Since this approach can only produce the topology of the plasma potential but not its absolute values, a specific scaling has to be  assumed. This is achieved, by adjusting the electric potential to be consistent with probe measurements, which were performed at distances of 6.3, 8.0 and 9.7 mm. The magnetic field configuration, obtained from Hall-probe measurements following the method of \textcite{kruger_reconstruction_2018}, can be found in a previous publication \cite{held_velocity_2020}.

Figure \ref{fig:plasma_pot} a) shows the reconstructed plasma potential assuming $\Phi \propto \Psi$. The arrows in the figure indicate the direction of the electric field, perpendicular to the magnetic field lines. This plasma potential topology is in good agreement with measurements of Rauch \etal and Mishra \etal \cite{rauch_plasma_2012, mishra_2d_2011}. 

Figure \ref{fig:plasma_pot}b shows the potential above the racetrack position ($r = \SI{13.5}{\milli\meter}$) as well as the electric field in axial direction $E_z$ calculated from the potential. This electric field can now be used to model the particle movement in $z$ direction as:

\begin{equation}
    \frac{\mathrm{d} v_z}{\mathrm{d} t} = \frac{e E_z}{m}
\end{equation}

\subsubsection{Forces in $\varphi$ direction and electron density}

During their movement in $z$-direction, the particles are accelerated in azimuthal direction. In contrast to the prior work from the literature \cite{lundin_cross-field_2008, poolcharuansin_more_2012, panjan_asymmetric_2014}, we do not consider a two stream instability or additional azimuthal electric field, but only the momentum transfer from electrons to ions via Coulomb collisions.  

The drag force acting on the ions $F_{drag}$ can be calculated as 
\begin{equation}
F_{drag} = \eta_{\perp} e^2 n_e v_e
\label{eq:drag_force}
\end{equation}

with the electron density $n_e$, the electron drift velocity in azimuthal direction $v_e$ and the cross $B$ resistivity $\eta_{\perp}$. For a highly ionized plasma, $\eta_{\perp}$ can be calculated as \cite{chen_introduction_2016}: 

\begin{equation}
 \eta_{\perp} = \frac{2 \pi e^2 \sqrt{m_e} }{(4 \pi \epsilon_0)^2 (k_B T_e)^{3/2}} \ln \Lambda
\label{eq:eta}
\end{equation}

with the electron mass $m_e$, the electron temperature $T_e = \SI{4.5}{\eV}$ and $\Lambda = 12 \pi n_e \lambda_D^3$ depending on the Debye length $\lambda_D$.

The electron drift velocity $v_e$ consists of the $\vec{E} \times \vec{B}$ drift, the curvature drift and the diamagnetic drift \cite{kruger_reconstruction_2018, chen_introduction_2016}:

\begin{align}
\label{eq:ve-drift}
    &{v}_{E\times B} = \frac{ \vec{E} \times  \vec{B}}{B^2} \\
    &{v}_{c} = -\frac{v_{||}^2}{\omega_c} \vec{b} \times (\vec{b}\cdot \nabla)\vec{b}\\
    &{v}_{dia} =T_e \frac{\nabla p \times  \vec{B}}{e n_e B^2}
\end{align}

with the electron cyclotron frequency $\omega_c =eB/m_e$, the unit vector $\vec{b} = \vec{B}/B$, and the electron velocities $v_{||}$ parallel and $v_\perp$ perpendicular to the magnetic field. These drift velocities, together with equations \ref{eq:drag_force} and \ref{eq:eta} can be used to calculate the drag force acting on the ions as they travel through the plasma, leading to acceleration in the azimuthal direction. However, an estimation for the electron density is required for the diamagnetic drift as well as the momentum transfer from electrons to ions.

\paragraph{Electron density:} The spatial dependence of the electron density was estimated from the discharge current, from Langmuir probe measurements as well as from optical measurements. The maximum of the electron density $n_e$ is expected to be located at the pre-sheath edge close to the target surface. This density can be derived from the measured discharge current $I$ and the Bohm velocity $v_{B} = \sqrt{k_B T_e/M}$ as
\begin{equation}
    n_e \approx \frac{2 I}{0.61 e v_B A}
\end{equation}
with the target surface area $A$ and the factor of two accounting for the difference between the average current density across the target surface and the larger local current density above the racetrack \cite{hecimovic_probing_2017, held_transient_2021}. Assuming a mix of titanium and argon ions with an average mass of $M = \SI{44}{u}$ and neglecting multiply charged ions as well as secondary electrons, we find a maximum density of $n_e  \approx  \SI{1.8e20}{\per\cubic\meter}$. The density was also measured using a Langmuir probe at distances of 6.3, 8.0 and 9.7 mm. The spatial distribution of the electron density in $z$ direction was further assumed to roughly follow the square root of the titanium ion emission, since this emission depends on the product of electron density $n_e$ and titanium ion density $n_{Ti}^+$. This relationship requires singly charged titanium ions to be the dominant ion species and a constant electron temperature, which is not the case. But the relationship can still be useful as an indication of the expected shape of electron density distribution. The titanium ion emission was obtained using Abel-inverted optical imaging, using a recently described setup \cite{held_ionization_2023, held_transient_2021}. Based on these three pieces of information  we approximate the spatial electron density profile as:
\begin{equation}
\label{eq:ne_guess}
n_\mathrm{e}(z) = \frac{\SI{2.47e20}{\per\cubic\meter}}{\exp(L_1/z)+1} \left( \exp(-z/L_2) + \exp(-z/L_3) \right) + \SI{2.5e19}{\per\cubic\meter} ~ .
\end{equation}
with $L_1 = \SI{0.1}{\milli\meter}$,  $L_2 = \SI{0.8}{\milli\meter}$ and $L_3 = \SI{4.5}{\milli\meter}$.

\begin{figure}[ht]
\centering
\includegraphics[scale=0.6]{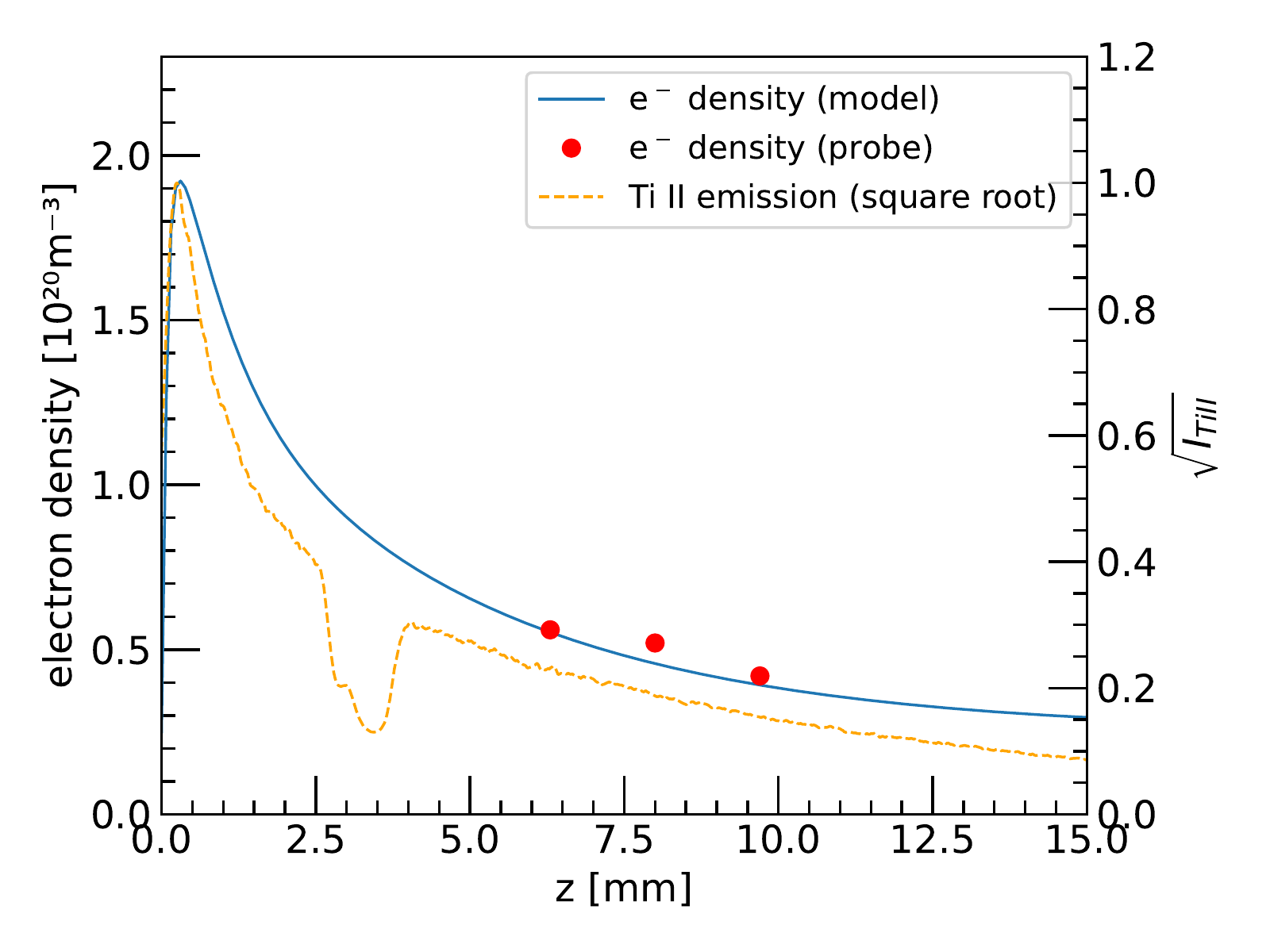}
\caption{Estimated electron density according to equation \ref{eq:ne_guess}, together with the results from probe measurements and the square root of Ti\,II emission.}
\label{fig:results:density}
\end{figure}

The resulting electron density is shown figure \ref{fig:results:density}, together with the probe measurements and the Abel inverted titanium ion emission, for comparison.

\begin{figure}[htb]
\centering
\includegraphics[width=75mm]{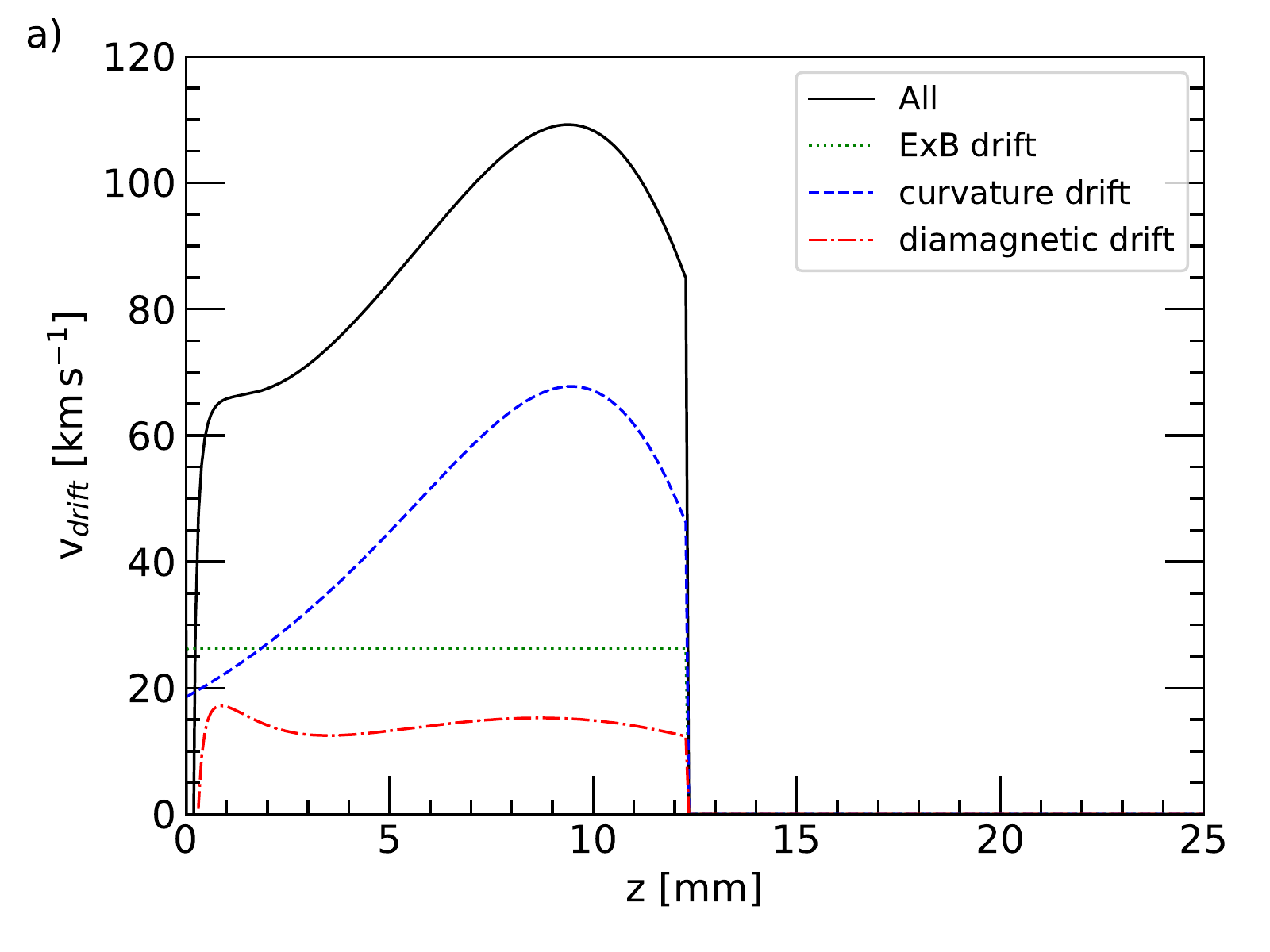}%
\includegraphics[width=75mm]{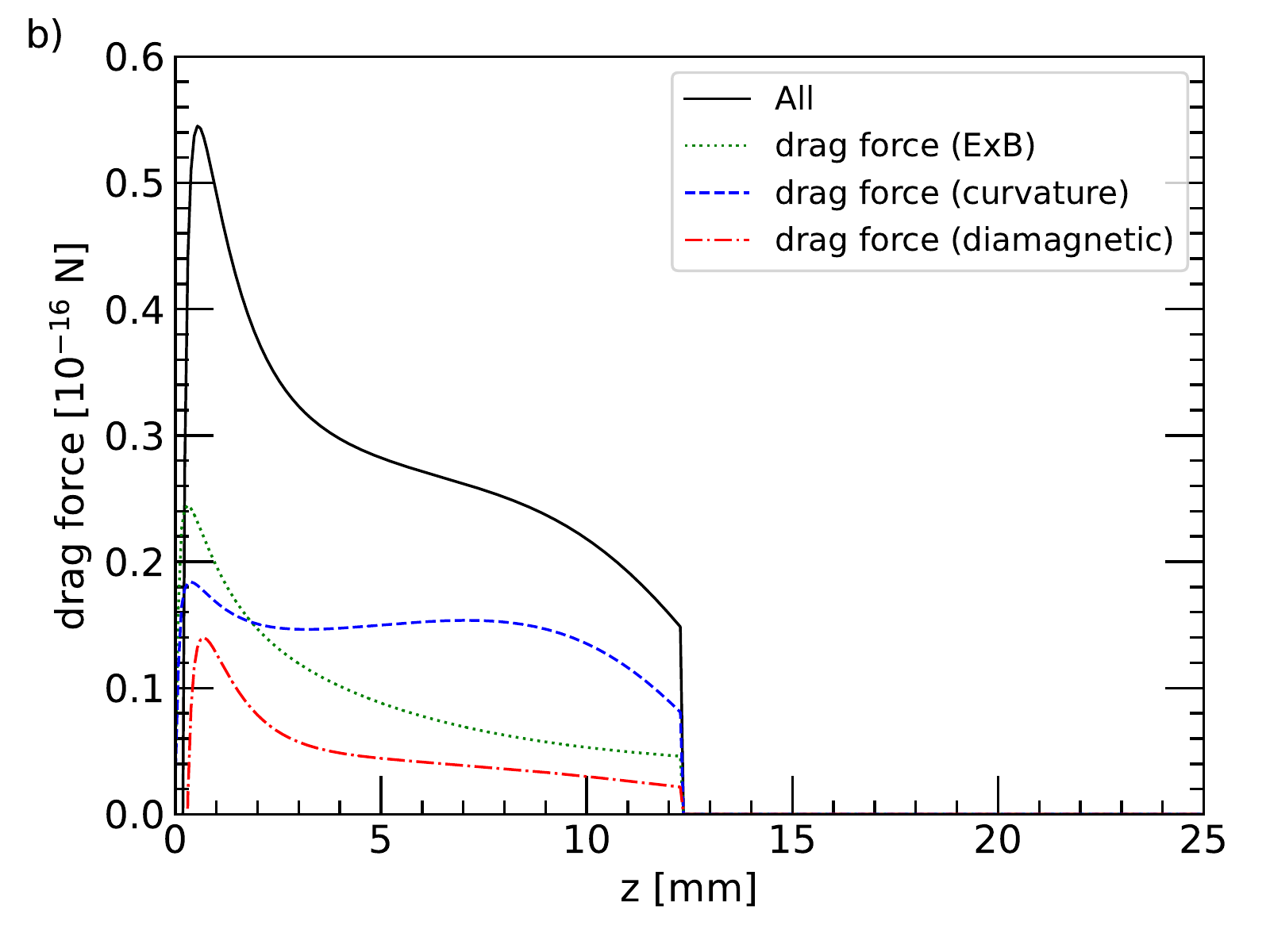}
\caption{a) Drift velocity of electrons. b) Drag force acting on the ions, caused by Coulomb collisions with the drifting electrons. }
\label{fig:drift_force}
\end{figure}

Based on the proposed electric field $\vec{E}(z)$, magnetic field $\vec{B}(z)$, and electron density $n_e(z)$, the different electron drift velocities as well as their sum can be calculated, as shown in figure \ref{fig:drift_force}a. Diamagnetic and \ExB drift are almost constant along $z$, which is a consequence of assuming the gradients for electron density and electric field to be similar as the gradients in the magnetic field configuration. For the diamagnetic drift and \ExB drift, we find drift velocities of about \SI{15}{\kilo\meter\per\second} and \SI{26}{\kilo\meter\per\second}, respectively. In contrast, the curvature drift velocity increases from about \SI{20}{\kilo\meter\per\second} close to the target to about \SI{68}{\kilo\meter\per\second} at \SI{10}{\milli\meter} and then begins to decrease. All drift velocities are set to zero for $z > \SI{12.3}{\milli\meter}$, because the gyroradius $r_L$ of the electrons becomes larger than half the gradient length scale of the magnetic field $B/\nabla B$, which we use as criterion for electron magnetization \cite{kruger_reconstruction_2018}. One can state that all drift mechanisms contribute similarly to the overall azimuthal electron drift velocity yielding values in the order of \SI{100}{\kilo\meter\per\second}, as expected \cite{rauch_estimating_2013, lundin_anomalous_2008}.

From the calculated drift velocities, we can now determine the azimuthal force acting on the ions, using equation \ref{eq:drag_force}.
Figure \ref{fig:drift_force}b shows this drag force and the individual contributions from the different drift velocities as a function of target distance. The drag force peaks close to the target surface due of the large electron density, but then decreases at larger target distances. 




\subsection{Test-particle simulation}
\label{sec:simu}

The model described above is solved using a test-particle Monte Carlo simulation (TPMC) with one spatial ($z$) and two velocity dimensions ($v_z$ and $v_{\varphi}$). This method was first used by \textcite{davis_monte_1960} and is comparable to a PIC simulation in which the fields are specified \textit{a priori} \cite{marchand_test-particle_2010}. The simulation considers an ensemble of $10^7$ particles that propagate in space according to their velocity and are accelerate according to the forces within small time steps $\Delta t = \SI{75}{\nano\second}$. Titanium ions are considered as the test particles and are introduced at $z = 0$, corresponding to ionization very close to the target surface \cite{held_ionization_2023}.

Particles are initialized with a Thompson distribution and are then accelerated by the electric field in $z$ direction and by the electron drag force in $\varphi$ direction. Particles are removed from the simulation if they either leave the simulation volume by moving beyond $z = \SI{40}{\milli\meter}$, or if they return to the target. In both cases, a new test particle is created at $z = 0$, to keep the total amount of particles constant. The simulation is performed over a time of \SI{60}{\micro\second}, ensuring full convergence to a steady state. Collisions are neglected in the simulation since the densities of the main collision partners are unknown. Further details on the simulation can be found in the appendix.

\begin{figure}[h]
\centering
\includegraphics[scale=0.5]{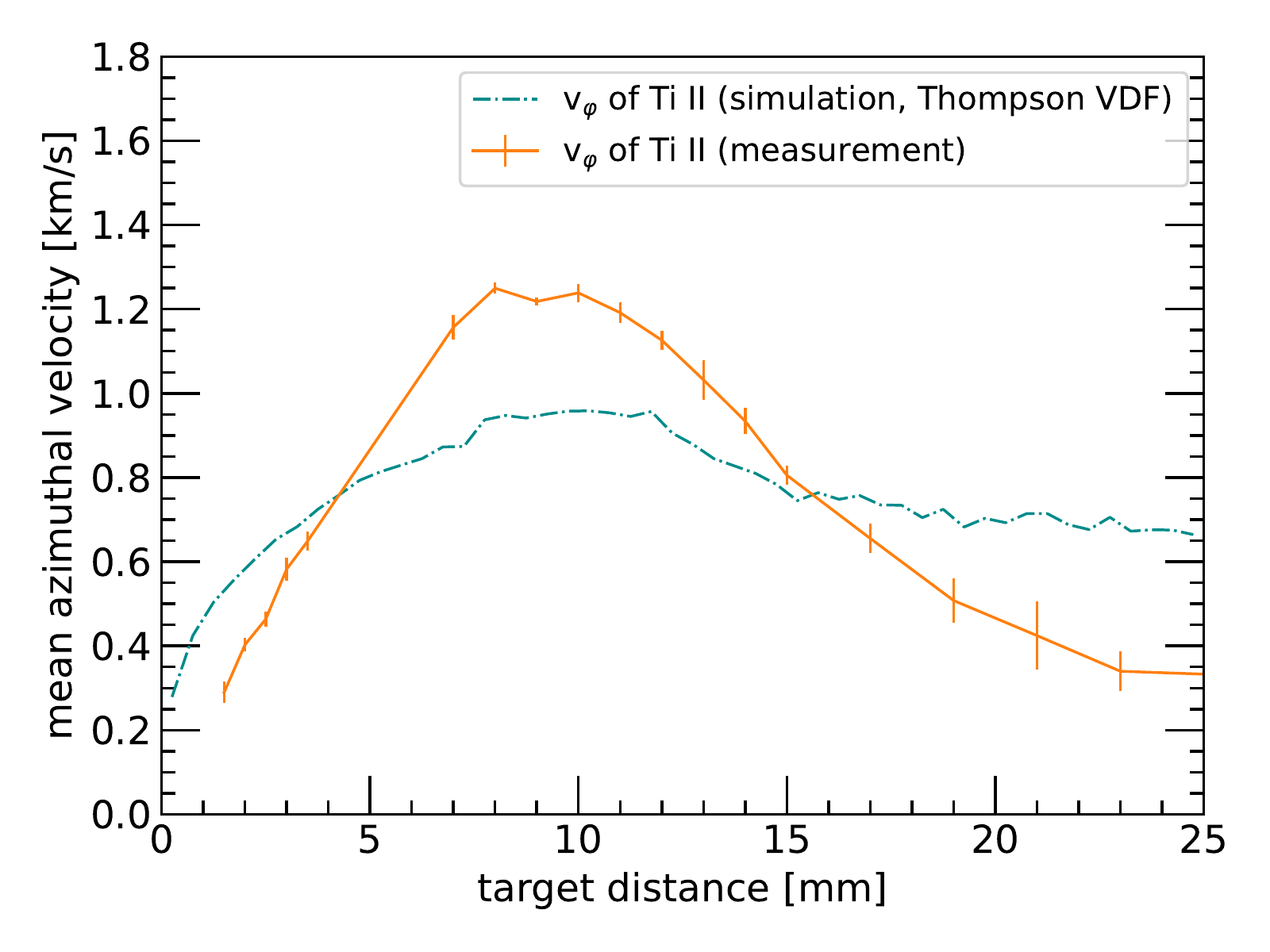}
\caption{Mean azimuthal velocity of titanium ion test particles together with the measurements presented in figure \ref{fig:results:vrot-zvar}. }
\label{fig:results:sim_vrot}
\end{figure}

The mean azimuthal velocity is extracted from the converged simulation results for $z$-positions from $z=0$ mm up to $z$ = 25  mm in steps of $\Delta z$ = 0.25 mm  by integrating the VDF of all particles within the defined interval [$z$, $z$+$\Delta z$]. A comparison of the simulation with the measurements is shown in figure \ref{fig:results:sim_vrot}. The simulation yields an increase of the azimuthal velocity in the vicinity of the target and a decrease at large distances from the target with a maximum at a distance of about 10 mm, in good agreement with the experiment. However, the simulation predicts a slightly smaller maximum velocity of only \SI{0.95}{\kilo\meter\per\second}, compared to the \SI{1.25}{\kilo\meter\per\second} found in the experiment. Furthermore, the simulation shows a much less steep decrease in azimuthal velocity for $z > \SI{12}{\milli\meter}$ than the experiment. 

For the difference in maximum velocity, we propose the influence of spokes as a possible reason. Since spokes posses strong azimuthal electric fields, they should be expected to additionally affect the azimuthal velocities. However, due to the complexity of spokes, an implementation of this influence within the simulation was not possible, since the wave phenomenon propagating in azimuthal direction breaks the symmetry of the 1d simulation. Based on the good agreement between simulation and experiment, the influence of spokes appears to be smaller than the drag force between electrons and ions, caused by Coulomb collisions. However, it should be noted, that this conclusion is not necessarily valid for all discharge conditions, since spokes under the present conditions have been observed to not be very strong \cite{held_ionization_2023}. 

The difference between simulation and experiment for $z > \SI{12}{\milli\meter}$ is likely caused by collisions. In the experiment, titanium ions at these target distances experience collisions with the background gas, which will lead them to slow down. Such collisions, however, are not included in the simulation, but a slight decrease in azimuthal velocity for $z > \SI{12}{\milli\meter}$ is nevertheless reproduced. In the absence of any forces, no change of azimuthal velocity should take place in this region. However, the velocity decrease is assigned to the filtering of particles by the electric field in $z$-direction: only particles with a certain minimum starting velocity can reach a certain $z$-distance. The higher these minimum starting velocity the smaller the transit time within the region of high azimuthal force. This consequently leads to a smaller degree of accumulation of azimuthal velocity for the particles which are able to reach larger distances. 

Despite these differences, the agreement between the simple simulation and the experiment is surprisingly good. Based on this agreement, we propose that at least a considerable part of the ion acceleration in azimuthal direction, observed in HiPIMS plasmas, is caused by the drag force from the drifting electrons on the ions via Coulomb collisions. This explanation differs from those found in the literature, which proposed different wave phenomena (the modified two stream instability or spokes) as the reason for the ion acceleration. It is likely, that these wave phenomena will also play a role in the azimuthal acceleration of ions. However, electron-ion collisions can clearly not be neglected.

\section{Conclusion}

The azimuthal velocity of titanium and argon ions was measured inside the magnetic trap region of a HiPIMS discharge. The velocity distribution function (VDF) of the ions was obtained using optical emission spectroscopy, thus allowing access to the space-resolved VDF inside the discharge, instead of only sampling ions that leave the plasma. 

Results showed the azimuthal ion velocity to increase with target distance, peaking at about \SI{1.55}{\kilo\meter\per\second} for argon ions and \SI{1.25}{\kilo\meter\per\second} for titanium ions. The difference between the maximum velocities was explained as partly caused by the difference in ion mass and partly by the different locations, where ionization occurs for the two species. Titanium neutrals were also found to follow the azimuthal ion movement of the ions, likely due to frequent charge exchange collisions between neutrals and ions.  

A model for the discharge was proposed, estimating the electric field and electron density inside the magnetic trap region from probe measurements and simple physical arguments. Based on this, electron drift velocities were calculated and the corresponding drag force, caused by Coulomb collisions between electrons and ions was obtained. A simple test particle simulation was performed to determine the azimuthal ion velocity under these conditions, only considering the drag force of electrons on the ions caused by collisions, and neglecting other aspects, such as spokes and ion-neutral collisions. The simulation showed surprisingly good agreement to the experiment, indicating that Coulomb collisions between the drifting electrons and the much slower ions might be the primary reason for the azimuthal ion movement in HiPIMS plasmas.  

The second part of this series of two papers will investigate how ions leaving the magnetic trap region are affected by the azimuthal drag force.

\section{Acknowledgments}

This work has been funded by the DFG within the framework of the collaborative research centre SFB-TR 87. 

\section*{Data availability}

The data that support the findings of this study are openly available at the following DOI: 10.5281/zenodo.7904947

\appendix

\section{Test particle Monte Carlo simulation}

After initializing the particle ensemble using defined starting conditions, a leapfrog algorithm is used to perform the particle movement. Accordingly the calculations within each timestep are as follows:

\begin{itemize}
    \item Calculating electric field and azimuthal drag force for current $z$-position 
    \item Accelerating for $\frac{\Delta t}{2}$ according to calculated electric field and azimuthal drag force ($v'_{z} = v_{z} + \frac{q \cdot E(z)}{m_{Ti}} \cdot \frac{\Delta t}{2} $ and $v'_{\varphi} = v_{\varphi} + \frac{F_{drag}(z)}{m_{Ti}} \cdot \frac{\Delta t}{2} $) 
    \item Moving in $z$-direction according to calculated $v'_z$-values ($z'' = z + v'_z \cdot \Delta t$) 
    \item Calculating electric field and azimuthal drag force for current $z$-position 
    \item Accelerating for $\frac{\Delta t}{2}$ according to calculated electric field and azimuthal drag force ($v''_{z} = v'_{z} + \frac{q \cdot E(z'')}{m_{Ti}} \cdot \frac{\Delta t}{2}$ and $v''_{\varphi} = v'_{\varphi} + \frac{F_{drag}(z'')}{m_{Ti}} \cdot \frac{\Delta t}{2}$) 
    \item Replace certain particles of the ensemble according to the boundary conditions
\end{itemize}

As initial condition the z-position of all particles is set to zero, while $v_z$ and $v_{\varphi}$ are selected randomly from a Thompson distribution. Within the selection process a 3-dimensional velocity distribution function is calculated according to a cosine angular and a Thompson energy distribution. Randomly selected values from the projections of this distribution in z- and $\varphi$ - direction are used for $v_z$ and $v_{\varphi}$. This initial condition is motivated by the expected angular and energy distribution of sputtered titanium neutrals, which are ionized in the vicinity of the target surface. 

The boundary conditions are as follows: particles reaching $z\leq$ 0 mm or $z\geq$ = 40 mm are removed from the ensemble, which mimics the loss of particles either to the target surface or to the substrate. To keep the total number of particles in the simulation constant, all removed particles are replaced with new particles at $z$ = 0 mm with $v_z$ and $v_{\varphi}$ being selected from the 3-dimensional Thompson distribution of sputtered titanium with its angular and energy dependencies being projected on the $z$- and $\varphi$ - direction. These initial conditions correspond to an ionisation of all sputtered particles directly after ejection from the target surface. A convergence of the simulation is reached, when the number of particles entering and leaving each volume element is equal and the distribution of particle densities and velocities reach steady state.

\newpage
\printbibliography

@thesis{held_transient_2021,
	title = {Transient transport phenomena in high power impulse magnetron sputtering discharges},
	rights = {http://hss-opus.ub.ruhr-uni-bochum.de/opus4/default/license/index/{licId}/3},
	url = {http://hss-opus.ub.ruhr-unibochum.de/opus4/frontdoor/index/index/docId/8326},
	type = {phdthesis},
	author = {Held, Julian},
	urldate = {2021-10-12},
	date = {2021-08-31},
	langid = {german},
}

@article{maas_synchronising_2021,
	title = {Synchronising optical emission spectroscopy to spokes in magnetron sputtering discharges},
	volume = {30},
	issn = {0963-0252, 1361-6595},
	url = {https://iopscience.iop.org/article/10.1088/1361-6595/ac3210},
	doi = {10.1088/1361-6595/ac3210},
	pages = {125006},
	number = {12},
	journaltitle = {Plasma Sources Science and Technology},
	shortjournal = {Plasma Sources Sci. Technol.},
	author = {Maaß, Philipp A and Schulz-von der Gathen, Volker and von Keudell, Achim and Held, Julian},
	urldate = {2022-03-08},
	date = {2021-12-01},
	langid = {english},
}

@article{biskup_influence_2018,
	title = {Influence of spokes on the ionized metal flux fraction in chromium high power impulse magnetron sputtering},
	volume = {51},
	issn = {0022-3727, 1361-6463},
	url = {http://stacks.iop.org/0022-3727/51/i=11/a=115201?key=crossref.cfe0515e3181a425317f84f885e1dc90},
	doi = {10.1088/1361-6463/aaac15},
	pages = {115201},
	number = {11},
	journaltitle = {Journal of Physics D: Applied Physics},
	author = {Biskup, B and Maszl, C and Breilmann, W and Held, J and Böke, M and Benedikt, J and von Keudell, A},
	urldate = {2018-03-24},
	date = {2018-03-21},
	langid = {english},
}

@article{breilmann_fast_2017,
	title = {Fast charge exchange ions in high power impulse magnetron sputtering of titanium as probes for the electrical potential},
	volume = {26},
	issn = {1361-6595},
	url = {http://stacks.iop.org/0963-0252/26/i=3/a=035007?key=crossref.619d0caace9bcfd6e5568cfae51bbe78},
	doi = {10.1088/1361-6595/aa56e5},
	pages = {035007},
	number = {3},
	journaltitle = {Plasma Sources Science and Technology},
	author = {Breilmann, W and Maszl, C and Keudell, A von},
	urldate = {2018-03-24},
	date = {2017-02-15},
	langid = {english},
}

@article{rauch_estimating_2013,
	title = {Estimating electron drift velocities in magnetron discharges},
	volume = {89},
	issn = {0042-207X},
	url = {http://www.sciencedirect.com/science/article/pii/S0042207X12004010},
	doi = {10.1016/j.vacuum.2012.09.002},
	series = {Including rapid communications, original articles and a special section with papers from the Seventeenth International Conference on Surface Modification of Materials by Ion Beams ({SMMIB}), 13–17 September 2011, Harbin, China},
	pages = {53--56},
	journaltitle = {Vacuum},
	shortjournal = {Vacuum},
	author = {Rauch, Albert and Anders, André},
	urldate = {2019-10-17},
	date = {2013-03-01},
	langid = {english},
}

@article{anders_drifting_2012,
	title = {Drifting localization of ionization runaway: Unraveling the nature of anomalous transport in high power impulse magnetron sputtering},
	volume = {111},
	issn = {0021-8979, 1089-7550},
	url = {http://aip.scitation.org/doi/10.1063/1.3692978},
	doi = {10.1063/1.3692978},
	shorttitle = {Drifting localization of ionization runaway},
	pages = {053304},
	number = {5},
	journaltitle = {Journal of Applied Physics},
	author = {Anders, André and Ni, Pavel and Rauch, Albert},
	urldate = {2018-03-23},
	date = {2012-03},
	langid = {english},
}

@article{kozyrev_optical_2011,
	title = {Optical studies of plasma inhomogeneities in a high-current pulsed magnetron discharge},
	volume = {37},
	issn = {1063-780X, 1562-6938},
	url = {http://link.springer.com/10.1134/S1063780X11060122},
	doi = {10.1134/S1063780X11060122},
	pages = {621--627},
	number = {7},
	journaltitle = {Plasma Physics Reports},
	author = {Kozyrev, A. V. and Sochugov, N. S. and Oskomov, K. V. and Zakharov, A. N. and Odivanova, A. N.},
	urldate = {2018-04-04},
	date = {2011-07},
	langid = {english},
}

@article{ehiasarian_high_2012,
	title = {High power impulse magnetron sputtering discharges: Instabilities and plasma self-organization},
	volume = {100},
	issn = {0003-6951, 1077-3118},
	url = {http://aip.scitation.org/doi/10.1063/1.3692172},
	doi = {10.1063/1.3692172},
	shorttitle = {High power impulse magnetron sputtering discharges},
	pages = {114101},
	number = {11},
	journaltitle = {Applied Physics Letters},
	author = {Ehiasarian, A. P. and Hecimovic, A. and de los Arcos, T. and New, R. and Schulz-von der Gathen, V. and Böke, M. and Winter, J.},
	urldate = {2018-04-08},
	date = {2012-03-12},
	langid = {english},
}

@article{panjan_plasma_2017,
	title = {Plasma potential of a moving ionization zone in {DC} magnetron           sputtering},
	volume = {121},
	issn = {0021-8979},
	url = {https://aip.scitation.org/doi/10.1063/1.4974944},
	doi = {10.1063/1.4974944},
	pages = {063302},
	number = {6},
	journaltitle = {Journal of Applied Physics},
	shortjournal = {Journal of Applied Physics},
	author = {Panjan, Matjaž and Anders, André},
	urldate = {2021-06-15},
	date = {2017-02-09},
}

@article{hecimovic_probing_2017,
	title = {Probing the electron density in {HiPIMS} plasmas by target inserts},
	volume = {50},
	issn = {0022-3727, 1361-6463},
	url = {https://iopscience.iop.org/article/10.1088/1361-6463/aa9914},
	doi = {10.1088/1361-6463/aa9914},
	pages = {505204},
	number = {50},
	journaltitle = {Journal of Physics D: Applied Physics},
	shortjournal = {J. Phys. D: Appl. Phys.},
	author = {Hecimovic, Ante and Held, Julian and Schulz-von der Gathen, Volker and Breilmann, Wolfgang and Maszl, Christian and von Keudell, Achim},
	urldate = {2021-06-15},
	date = {2017-12-20},
}

@article{held_velocity_2020,
	title = {Velocity distribution of metal ions in the target region of {HiPIMS}: the role of Coulomb collisions},
	volume = {29},
	issn = {1361-6595},
	url = {https://iopscience.iop.org/article/10.1088/1361-6595/abbf94},
	doi = {10.1088/1361-6595/abbf94},
	shorttitle = {Velocity distribution of metal ions in the target region of {HiPIMS}},
	pages = {125003},
	number = {12},
	journaltitle = {Plasma Sources Science and Technology},
	shortjournal = {Plasma Sources Sci. Technol.},
	author = {Held, J and Thiemann-Monjé, S and von Keudell, A and Schulz-von der Gathen, V},
	urldate = {2021-06-15},
	date = {2020-12-09},
}

@article{held_pattern_2020,
	title = {Pattern Formation in High Power Impulse Magnetron Sputtering ({HiPIMS}) Plasmas},
	volume = {40},
	issn = {1572-8986},
	url = {https://doi.org/10.1007/s11090-019-10052-3},
	doi = {10.1007/s11090-019-10052-3},
	pages = {643--660},
	number = {3},
	journaltitle = {Plasma Chemistry and Plasma Processing},
	shortjournal = {Plasma Chem Plasma Process},
	author = {Held, Julian and von Keudell, Achim},
	urldate = {2021-06-15},
	date = {2020-05-01},
	langid = {english},
}

@article{held_electron_2020,
	title = {Electron density, temperature and the potential structure of spokes in {HiPIMS}},
	volume = {29},
	issn = {1361-6595},
	url = {https://iopscience.iop.org/article/10.1088/1361-6595/ab5e46},
	doi = {10.1088/1361-6595/ab5e46},
	pages = {025006},
	number = {2},
	journaltitle = {Plasma Sources Science and Technology},
	shortjournal = {Plasma Sources Sci. Technol.},
	author = {Held, J and Maaß, P A and Schulz-von der Gathen, V and von Keudell, A},
	urldate = {2021-06-15},
	date = {2020-02-06},
}

@article{hnilica_effect_2018,
	title = {Effect of magnetic field on spoke behaviour in {HiPIMS} plasma},
	volume = {51},
	issn = {0022-3727, 1361-6463},
	url = {https://iopscience.iop.org/article/10.1088/1361-6463/aaa7d3},
	doi = {10.1088/1361-6463/aaa7d3},
	pages = {095204},
	number = {9},
	journaltitle = {Journal of Physics D: Applied Physics},
	shortjournal = {J. Phys. D: Appl. Phys.},
	author = {Hnilica, J and Klein, P and Šlapanská, M and Fekete, M and Vašina, P},
	urldate = {2021-06-15},
	date = {2018-03-07},
}

@article{alami_ion-assisted_2005,
	title = {Ion-assisted physical vapor deposition for enhanced film properties on nonflat surfaces},
	volume = {23},
	issn = {0734-2101},
	url = {http://avs.scitation.org/doi/abs/10.1116/1.1861049},
	doi = {10.1116/1.1861049},
	pages = {278--280},
	number = {2},
	journaltitle = {Journal of Vacuum Science \& Technology A: Vacuum, Surfaces, and Films},
	shortjournal = {Journal of Vacuum Science \& Technology A: Vacuum, Surfaces, and Films},
	author = {Alami, J. and Persson, P. O. Å. and Music, D. and Gudmundsson, J. T. and Bohlmark, J. and Helmersson, U.},
	date = {2005-02-10},
}

@article{anders_recycling_2012,
	title = {The ‘recycling trap’: a generalized explanation of discharge runaway in high-power impulse magnetron sputtering},
	volume = {45},
	issn = {0022-3727},
	url = {http://stacks.iop.org/0022-3727/45/i=1/a=012003},
	doi = {10.1088/0022-3727/45/1/012003},
	shorttitle = {The ‘recycling trap’},
	pages = {012003},
	number = {1},
	journaltitle = {Journal of Physics D: Applied Physics},
	shortjournal = {J. Phys. D: Appl. Phys.},
	author = {Anders, A. and Čapek, J. and Hála, M. and Martinu, L.},
	urldate = {2018-02-26},
	date = {2012},
	langid = {english},
}

@article{anders_deposition_2010,
	title = {Deposition rates of high power impulse magnetron sputtering: Physics and economics},
	volume = {28},
	issn = {0734-2101},
	url = {https://avs.scitation.org/doi/abs/10.1116/1.3299267},
	doi = {10.1116/1.3299267},
	shorttitle = {Deposition rates of high power impulse magnetron sputtering},
	pages = {783--790},
	number = {4},
	journaltitle = {Journal of Vacuum Science \& Technology A},
	shortjournal = {Journal of Vacuum Science \& Technology A},
	author = {Anders, André},
	urldate = {2018-09-12},
	date = {2010-06-29},
}

@article{anders_drifting_2013,
	title = {Drifting potential humps in ionization zones: The “propeller blades” of high power impulse magnetron sputtering},
	volume = {103},
	issn = {0003-6951},
	url = {http://aip.scitation.org/doi/full/10.1063/1.4823827},
	doi = {10.1063/1.4823827},
	shorttitle = {Drifting potential humps in ionization zones},
	pages = {144103},
	number = {14},
	journaltitle = {Applied Physics Letters},
	shortjournal = {Appl. Phys. Lett.},
	author = {Anders, André and Panjan, P and Franz, Robert and Andersson, Joakim and Ni, Pavel},
	urldate = {2017-02-10},
	date = {2013-09-30},
}

@article{bohlmark_ionization_2005,
	title = {Ionization of sputtered metals in high power pulsed magnetron sputtering},
	volume = {23},
	issn = {0734-2101, 1520-8559},
	url = {http://scitation.aip.org/content/avs/journal/jvsta/23/1/10.1116/1.1818135},
	doi = {10.1116/1.1818135},
	pages = {18--22},
	number = {1},
	journaltitle = {Journal of Vacuum Science \& Technology A},
	author = {Bohlmark, Johan and Alami, Jones and Christou, Chris and Ehiasarian, Arutiun P. and Helmersson, Ulf},
	urldate = {2016-12-06},
	date = {2005-01-01},
}

@article{brenning_understanding_2012,
	title = {Understanding deposition rate loss in high power impulse magnetron sputtering: I. Ionization-driven electric fields},
	volume = {21},
	issn = {0963-0252},
	url = {http://stacks.iop.org/0963-0252/21/i=2/a=025005},
	doi = {10.1088/0963-0252/21/2/025005},
	shorttitle = {Understanding deposition rate loss in high power impulse magnetron sputtering},
	pages = {025005},
	number = {2},
	journaltitle = {Plasma Sources Science and Technology},
	shortjournal = {Plasma Sources Sci. Technol.},
	author = {Brenning, N. and Huo, C. and Lundin, D. and Raadu, M. A. and Vitelaru, C. and Stancu, G. D. and Minea, T. and Helmersson, U.},
	urldate = {2016-12-13},
	date = {2012},
	langid = {english},
}

@article{brinkmann_axisymmetric_2020,
	title = {Axisymmetric magnetically enhanced discharges described in terms of flux coordinates},
	volume = {27},
	issn = {1070-664X},
	url = {https://aip.scitation.org/doi/full/10.1063/1.5140320},
	doi = {10.1063/1.5140320},
	pages = {053504},
	number = {5},
	journaltitle = {Physics of Plasmas},
	shortjournal = {Physics of Plasmas},
	author = {Brinkmann, Ralf Peter and Krüger, Dennis},
	urldate = {2020-05-21},
	date = {2020-05-01},
	note = {Publisher: American Institute of Physics},
}

@article{gudmundsson_spatial_2002,
	title = {Spatial and temporal behavior of the plasma parameters in a pulsed magnetron discharge},
	volume = {161},
	issn = {0257-8972},
	url = {http://www.sciencedirect.com/science/article/pii/S0257897202005182},
	doi = {10.1016/S0257-8972(02)00518-2},
	pages = {249--256},
	number = {2},
	journaltitle = {Surface and Coatings Technology},
	shortjournal = {Surface and Coatings Technology},
	author = {Gudmundsson, J. T. and Alami, J. and Helmersson, U.},
	urldate = {2016-05-09},
	date = {2002},
}

@article{hecimovic_spokes_2018,
	title = {Spokes in high power impulse magnetron sputtering plasmas},
	volume = {51},
	issn = {0022-3727},
	url = {https://doi.org/10.1088%2F1361-6463%2Faadaa1},
	doi = {10.1088/1361-6463/aadaa1},
	pages = {453001},
	number = {45},
	journaltitle = {Journal of Physics D: Applied Physics},
	shortjournal = {J. Phys. D: Appl. Phys.},
	author = {Hecimovic, Ante and von Keudell, Achim},
	urldate = {2020-06-30},
	date = {2018-09},
	langid = {english},
	note = {Publisher: {IOP} Publishing},
}

@article{hnilica_revisiting_2020,
	title = {Revisiting particle dynamics in {HiPIMS} discharges. I. General effects},
	volume = {128},
	issn = {0021-8979},
	url = {https://aip.scitation.org/doi/full/10.1063/5.0009378},
	doi = {10.1063/5.0009378},
	pages = {043303},
	number = {4},
	journaltitle = {Journal of Applied Physics},
	shortjournal = {Journal of Applied Physics},
	author = {Hnilica, Jaroslav and Klein, Peter and Vašina, Petr and Snyders, Rony and Britun, Nikolay},
	urldate = {2020-08-03},
	date = {2020-07-28},
	note = {Publisher: American Institute of Physics},
}

@article{hnilica_revisiting_2020-1,
	title = {Revisiting particle dynamics in {HiPIMS} discharges. {II}. Plasma pulse effects},
	volume = {128},
	issn = {0021-8979},
	url = {https://aip.scitation.org/doi/full/10.1063/5.0009380},
	doi = {10.1063/5.0009380},
	pages = {043304},
	number = {4},
	journaltitle = {Journal of Applied Physics},
	shortjournal = {Journal of Applied Physics},
	author = {Hnilica, Jaroslav and Klein, Peter and Vašina, Petr and Snyders, Rony and Britun, Nikolay},
	urldate = {2020-08-03},
	date = {2020-07-28},
	note = {Publisher: American Institute of Physics},
}

@article{horwat_spatial_2008,
	title = {Spatial distribution of average charge state and deposition rate in high power impulse magnetron sputtering of copper},
	volume = {41},
	issn = {0022-3727},
	url = {http://stacks.iop.org/0022-3727/41/i=13/a=135210},
	doi = {10.1088/0022-3727/41/13/135210},
	pages = {135210},
	number = {13},
	journaltitle = {Journal of Physics D: Applied Physics},
	shortjournal = {J. Phys. D: Appl. Phys.},
	author = {Horwat, David and Anders, André},
	urldate = {2017-01-10},
	date = {2008},
	langid = {english},
}

@article{kanitz_two_2016,
	title = {Two dimensional spatial Argon metastable dynamics in {HiPIMS} discharges},
	volume = {49},
	issn = {0022-3727},
	url = {http://stacks.iop.org/0022-3727/49/i=12/a=125203},
	doi = {10.1088/0022-3727/49/12/125203},
	pages = {125203},
	number = {12},
	journaltitle = {Journal of Physics D: Applied Physics},
	shortjournal = {J. Phys. D: Appl. Phys.},
	author = {Kanitz, Alexander and Hecimovic, Ante and Böke, Marc and Winter, Jörg},
	urldate = {2016-08-17},
	date = {2016},
	langid = {english},
}

@article{kouznetsov_novel_1999,
	title = {A novel pulsed magnetron sputter technique utilizing very high target power densities},
	volume = {122},
	issn = {0257-8972},
	url = {https://www.sciencedirect.com/science/article/pii/S0257897299002923},
	doi = {10.1016/S0257-8972(99)00292-3},
	pages = {290--293},
	number = {2},
	journaltitle = {Surface and Coatings Technology},
	shortjournal = {Surface and Coatings Technology},
	author = {Kouznetsov, Vladimir and Macák, Karol and Schneider, Jochen M. and Helmersson, Ulf and Petrov, Ivan},
	urldate = {2021-06-17},
	date = {1999-12-15},
	langid = {english},
}

@article{kruger_reconstruction_2018,
	title = {Reconstruction of the static magnetic field of a magnetron},
	volume = {25},
	issn = {1070-664X},
	url = {https://aip.scitation.org/doi/abs/10.1063/1.5024983},
	doi = {10.1063/1.5024983},
	pages = {061207},
	number = {6},
	journaltitle = {Physics of Plasmas},
	shortjournal = {Physics of Plasmas},
	author = {Krüger, Dennis and Köhn, Kevin and Gallian, Sara and Brinkmann, Ralf Peter},
	urldate = {2019-10-17},
	date = {2018-05-22},
}

@article{lockwood_estrin_triple_2017,
	title = {Triple probe interrogation of spokes in a {HiPIMS} discharge},
	volume = {50},
	issn = {0022-3727},
	url = {http://stacks.iop.org/0022-3727/50/i=29/a=295201},
	doi = {10.1088/1361-6463/aa7544},
	pages = {295201},
	number = {29},
	journaltitle = {Journal of Physics D: Applied Physics},
	shortjournal = {J. Phys. D: Appl. Phys.},
	author = {Lockwood Estrin, F. and Karkari, S. K. and Bradley, J. W.},
	date = {2017},
	langid = {english},
}

@article{lundin_anomalous_2008,
	title = {Anomalous electron transport in high power impulse magnetron sputtering},
	volume = {17},
	issn = {0963-0252},
	url = {https://doi.org/10.1088%2F0963-0252%2F17%2F2%2F025007},
	doi = {10.1088/0963-0252/17/2/025007},
	pages = {025007},
	number = {2},
	journaltitle = {Plasma Sources Science and Technology},
	shortjournal = {Plasma Sources Sci. Technol.},
	author = {Lundin, Daniel and Helmersson, Ulf and Kirkpatrick, Scott and Rohde, Suzanne and Brenning, Nils},
	urldate = {2020-05-28},
	date = {2008-03},
	langid = {english},
	note = {Publisher: {IOP} Publishing},
}

@article{lundin_cross-field_2008,
	title = {Cross-field ion transport during high power impulse magnetron sputtering},
	volume = {17},
	issn = {0963-0252},
	url = {http://stacks.iop.org/0963-0252/17/i=3/a=035021},
	doi = {10.1088/0963-0252/17/3/035021},
	pages = {035021},
	number = {3},
	journaltitle = {Plasma Sources Science and Technology},
	shortjournal = {Plasma Sources Sci. Technol.},
	author = {Lundin, Daniel and Larsson, Petter and Wallin, Erik and Lattemann, Martina and Brenning, Nils and Helmersson, Ulf},
	urldate = {2018-04-02},
	date = {2008},
	langid = {english},
}

@article{mishra_evolution_2010,
	title = {The evolution of the plasma potential in a {HiPIMS} discharge and its relationship to deposition rate},
	volume = {19},
	issn = {0963-0252},
	url = {http://stacks.iop.org/0963-0252/19/i=4/a=045014},
	doi = {10.1088/0963-0252/19/4/045014},
	pages = {045014},
	number = {4},
	journaltitle = {Plasma Sources Science and Technology},
	shortjournal = {Plasma Sources Sci. Technol.},
	author = {Mishra, Anurag and Kelly, P. J. and Bradley, J. W.},
	urldate = {2017-01-14},
	date = {2010},
	langid = {english},
}

@article{panjan_asymmetric_2014,
	title = {Asymmetric particle fluxes from drifting ionization zones in sputtering magnetrons},
	volume = {23},
	issn = {0963-0252},
	url = {http://stacks.iop.org/0963-0252/23/i=2/a=025007},
	doi = {10.1088/0963-0252/23/2/025007},
	pages = {025007},
	number = {2},
	journaltitle = {Plasma Sources Science and Technology},
	shortjournal = {Plasma Sources Sci. Technol.},
	author = {Panjan, Matjaž and Franz, Robert and Anders, André},
	urldate = {2018-08-07},
	date = {2014},
	langid = {english},
}

@article{poolcharuansin_ionized_2012,
	title = {Ionized metal flux fraction measurements in {HiPIMS} discharges},
	volume = {45},
	issn = {0022-3727},
	url = {http://stacks.iop.org/0022-3727/45/i=32/a=322001},
	doi = {10.1088/0022-3727/45/32/322001},
	pages = {322001},
	number = {32},
	journaltitle = {Journal of Physics D: Applied Physics},
	shortjournal = {J. Phys. D: Appl. Phys.},
	author = {Poolcharuansin, P. and Bowes, M. and Petty, T. J. and Bradley, J. W.},
	urldate = {2017-12-11},
	date = {2012},
	langid = {english},
}

@article{poolcharuansin_more_2012,
	title = {More evidence for azimuthal ion spin in {HiPIMS} discharges},
	volume = {21},
	issn = {0963-0252},
	url = {http://stacks.iop.org/0963-0252/21/i=1/a=015001},
	doi = {10.1088/0963-0252/21/1/015001},
	pages = {015001},
	number = {1},
	journaltitle = {Plasma Sources Science and Technology},
	shortjournal = {Plasma Sources Sci. Technol.},
	author = {Poolcharuansin, P. and Liebig, B. and Bradley, J. W.},
	urldate = {2018-03-14},
	date = {2012},
	langid = {english},
}

@article{rauch_plasma_2012,
	title = {Plasma potential mapping of high power impulse magnetron sputtering discharges},
	volume = {111},
	issn = {0021-8979},
	url = {https://aip.scitation.org/doi/abs/10.1063/1.3700242},
	doi = {10.1063/1.3700242},
	pages = {083302},
	number = {8},
	journaltitle = {Journal of Applied Physics},
	shortjournal = {Journal of Applied Physics},
	author = {Rauch, Albert and Mendelsberg, Rueben J. and Sanders, Jason M. and Anders, André},
	urldate = {2019-03-22},
	date = {2012-04-15},
}

@article{sarakinos_process_2007,
	title = {Process characteristics and film properties upon growth of {TiO} x films by high power pulsed magnetron sputtering},
	volume = {40},
	issn = {0022-3727},
	url = {http://stacks.iop.org/0022-3727/40/i=7/a=037},
	doi = {10.1088/0022-3727/40/7/037},
	pages = {2108},
	number = {7},
	journaltitle = {Journal of Physics D: Applied Physics},
	shortjournal = {J. Phys. D: Appl. Phys.},
	author = {Sarakinos, K. and Alami, J. and Wuttig, M.},
	urldate = {2017-01-31},
	date = {2007},
	langid = {english},
}

@article{vitelaru_argon_2012,
	title = {Argon metastables in {HiPIMS}: time-resolved tunable diode-laser diagnostics},
	volume = {21},
	issn = {0963-0252},
	url = {http://stacks.iop.org/0963-0252/21/i=2/a=025010},
	doi = {10.1088/0963-0252/21/2/025010},
	shorttitle = {Argon metastables in {HiPIMS}},
	pages = {025010},
	number = {2},
	journaltitle = {Plasma Sources Science and Technology},
	shortjournal = {Plasma Sources Sci. Technol.},
	author = {Vitelaru, C. and Lundin, D. and Stancu, G. D. and Brenning, N. and Bretagne, J. and Minea, T.},
	urldate = {2016-04-20},
	date = {2012},
	langid = {english},
}

@article{held_velocity_2018,
	title = {Velocity distribution of titanium neutrals in the target region of high power impulse magnetron sputtering discharges},
	volume = {27},
	issn = {0963-0252},
	url = {https://doi.org/10.1088%2F1361-6595%2Faae236},
	doi = {10.1088/1361-6595/aae236},
	pages = {105012},
	number = {10},
	journaltitle = {Plasma Sources Science and Technology},
	shortjournal = {Plasma Sources Sci. Technol.},
	author = {Held, J. and Hecimovic, A. and von Keudell, Achim and Schulz-von der Gathen, Volker},
	urldate = {2020-06-30},
	date = {2018-10},
	langid = {english},
	note = {Publisher: {IOP} Publishing},
}

@article{held_spoke-resolved_2022,
	title = {Spoke-resolved electron density, temperature and potential in direct current magnetron sputtering and {HiPIMS} discharges},
	issn = {0963-0252},
	url = {http://iopscience.iop.org/article/10.1088/1361-6595/ac87ce},
	doi = {10.1088/1361-6595/ac87ce},
	journaltitle = {Plasma Sources Science and Technology},
	shortjournal = {Plasma Sources Sci. Technol.},
	author = {Held, Julian and George, Mathews and von Keudell, Achim},
	urldate = {2022-08-09},
	date = {2022},
	langid = {english},
}

@article{hagelaar_two-dimensional_2002,
	title = {Two-dimensional model of a stationary plasma thruster},
	volume = {91},
	issn = {0021-8979, 1089-7550},
	url = {http://aip.scitation.org/doi/10.1063/1.1465125},
	doi = {10.1063/1.1465125},
	pages = {5592--5598},
	number = {9},
	journaltitle = {Journal of Applied Physics},
	shortjournal = {Journal of Applied Physics},
	author = {Hagelaar, G. J. M. and Bareilles, J. and Garrigues, L. and Boeuf, J. -P.},
	urldate = {2021-07-11},
	date = {2002-05},
	langid = {english},
}

@article{mishra_2d_2011,
	title = {The 2D plasma potential distribution in a {HiPIMS} discharge},
	volume = {44},
	issn = {0022-3727},
	url = {https://dx.doi.org/10.1088/0022-3727/44/42/425201},
	doi = {10.1088/0022-3727/44/42/425201},
	pages = {425201},
	number = {42},
	journaltitle = {Journal of Physics D: Applied Physics},
	shortjournal = {J. Phys. D: Appl. Phys.},
	author = {Mishra, A. and Kelly, P. J. and Bradley, J. W.},
	urldate = {2023-03-28},
	date = {2011-10},
	langid = {english},
}

@article{saloman_energy_2012,
	title = {Energy Levels and Observed Spectral Lines of Neutral and Singly Ionized Chromium, Cr I and Cr {II}},
	volume = {41},
	issn = {0047-2689, 1529-7845},
	url = {http://aip.scitation.org/doi/10.1063/1.4754694},
	doi = {10.1063/1.4754694},
	pages = {043103},
	number = {4},
	journaltitle = {Journal of Physical and Chemical Reference Data},
	shortjournal = {Journal of Physical and Chemical Reference Data},
	author = {Saloman, E. B.},
	urldate = {2023-03-28},
	date = {2012-12},
	langid = {english},
}

@book{chen_introduction_2016,
	location = {Cham Heidelberg New York Dordrecht London},
	edition = {Third edition},
	title = {Introduction to plasma physics and controlled fusion},
	isbn = {978-3-319-22309-4 978-3-319-22308-7},
	pagetotal = {490},
	publisher = {Springer},
	author = {Chen, Francis F.},
	date = {2016},
	note = {{OCLC}: 933265525},
}

@article{held_ionization_2023,
	title = {Ionization of sputtered material in high power impulse magnetron sputtering plasmas - comparison of titanium, chromium and aluminum},
	volume = {submitted},
	journaltitle = {Plasma Sources Science and Technology},
	author = {Held, Julian and Schulz-von der Gathen, Volker and von Keudell, Achim},
	date = {2023},
}

@article{russell_report_1929,
	title = {Report on Notation for Atomic Spectra},
	volume = {33},
	url = {https://link.aps.org/doi/10.1103/PhysRev.33.900},
	doi = {10.1103/PhysRev.33.900},
	pages = {900--906},
	number = {6},
	journaltitle = {Physical Review},
	shortjournal = {Phys. Rev.},
	author = {Russell, H. N. and Shenstone, A. G. and Turner, Louis A.},
	urldate = {2018-01-16},
	date = {1929-06-01},
}

@article{whaling_argon_1995,
	title = {Argon ion linelist and level energies in the hollow-cathode discharge},
	volume = {53},
	issn = {0022-4073},
	url = {https://www.sciencedirect.com/science/article/pii/002240739400102D},
	doi = {10.1016/0022-4073(94)00102-D},
	pages = {1--22},
	number = {1},
	journaltitle = {Journal of Quantitative Spectroscopy and Radiative Transfer},
	shortjournal = {Journal of Quantitative Spectroscopy and Radiative Transfer},
	author = {Whaling, W. and Anderson, W. H. C. and Carle, M. T. and Brault, J. W. and Zarem, H. A.},
	urldate = {2023-05-02},
	date = {1995-01-01},
	langid = {english},
}

@misc{kramida_nist_2022,
	title = {{NIST} Atomic Spectra Database},
	url = {https://doi.org/10.18434/T4W30F},
	author = {Kramida, A. and Ralchenko, Yu. and Reader, J. and {NIST ASD Team}},
	urldate = {2023-02-28},
	date = {2022},
	note = {Published: {NIST} Atomic Spectra Database (ver. 5.10), [Online]. Available: {\textbackslash}tthttps://physics.nist.gov/asd [2023, February 28]. National Institute of Standards and Technology, Gaithersburg, {MD}.},
}

@article{smirnov_tables_2000,
	title = {Tables for Cross Sections of the Resonant Charge Exchange Process},
	volume = {61},
	issn = {1402-4896},
	url = {https://iopscience.iop.org/article/10.1238/Physica.Regular.061a00595/meta},
	doi = {10.1238/Physica.Regular.061a00595},
	pages = {595},
	number = {5},
	journaltitle = {Physica Scripta},
	shortjournal = {Phys. Scr.},
	author = {Smirnov, B. M.},
	urldate = {2022-03-18},
	date = {2000-05-01},
	langid = {english},
	note = {Publisher: {IOP} Publishing},
}

@article{davis_monte_1960,
	title = {Monte Carlo Calculation of Molecular Flow Rates through a Cylindrical Elbow and Pipes of Other Shapes},
	volume = {31},
	issn = {0021-8979},
	url = {https://doi.org/10.1063/1.1735797},
	doi = {10.1063/1.1735797},
	pages = {1169--1176},
	number = {7},
	journaltitle = {Journal of Applied Physics},
	shortjournal = {Journal of Applied Physics},
	author = {Davis, D. H.},
	urldate = {2023-05-02},
	date = {1960},
}

@article{marchand_test-particle_2010,
	title = {Test-particle Simulation of Space Plasmas},
	volume = {8},
	url = {https://ui.adsabs.harvard.edu/abs/2010CCoPh...8..471M},
	doi = {10.4208/cicp.201009.280110a},
	pages = {471--483},
	journaltitle = {Communications in Computational Physics},
	author = {Marchand, Richard},
	urldate = {2023-05-02},
	date = {2010-06-01},
	note = {{ADS} Bibcode: 2010CCoPh...8..471M},
}

@collection{glocker_handbook_1995,
	location = {Bristol, {UK} ; Philadelphia},
	title = {Handbook of thin film process technology},
	isbn = {978-0-7503-0311-8 978-0-7503-0529-7},
	pagetotal = {1},
	publisher = {Institute of Physics Pub},
	editor = {Glocker, David A. and Shah, S. Ismat and Westwood, William D.},
	date = {1995},
	note = {{OCLC}: 33093216},
}

@article{schuttler_azimuthal_2023,
	title = {Azimuthal ion movement in {HiPIMS} plasmas - Part {II}: lateral growth fluxes},
	journaltitle = {(submitted)},
	author = {Schüttler, Steffen and Thiemann-Monje, Sascha and Held, Julian and von Keudell, Achim},
	date = {2023},
}

\end{document}